\documentclass[fleqn,usenatbib]{mnras}

\usepackage[T1]{fontenc}
\usepackage{ae,aecompl}

\usepackage{graphicx}    
\usepackage{amsmath}    
\usepackage{amssymb}    
\usepackage{stfloats}
\usepackage{pdflscape} 

\newcommand{\HI}{H\,\textsc{i}}
\newcommand{\Lya}{Ly$\alpha$}
\newcommand{\Lyb}{Ly$\beta$}

\title[Identifying \Lya\ systems with a CNN]{Harvesting the \Lya\ forest
with convolutional neural networks}
\author[Ting-Yun Cheng et al.]{Ting-Yun~Cheng,$^{1}$\thanks{E-mail:ting-yun.cheng@durham.ac.uk}
Ryan~J.~Cooke,$^{1}$
Gwen~Rudie$^{2}$
\\ \\
$^{1}$ Centre for Extragalactic Astronomy, Durham University, South Road, Durham DH1 3LE, UK\\
$^{2}$ The Observatories of the Carnegie Institution for Science, 813 Santa Barbara Street, Pasadena, CA 91101, USA\\
}

\date{Accepted XXX. Received YYY; in original form ZZZ}

\pubyear{2022}

\begin{document}
\label{firstpage}
\pagerange{\pageref{firstpage}--\pageref{lastpage}}
\maketitle

\begin{abstract}
We develop a machine learning based algorithm using a convolutional neural network (CNN) to identify low \HI\ column density \Lya\ absorption systems ($\log{N_{\mathrm{HI}}}/{\rm cm}^{-2}<17$) in the \Lya\ forest, and predict their physical properties, such as their \HI\ column density ($\log{N}_{\mathrm{HI}}/{\rm cm}^{-2}$), redshift ($z_{\mathrm{HI}}$), and Doppler width ($b_{\mathrm{HI}}$). Our CNN models are trained using simulated spectra (S/N $\simeq10$), and we test their performance on high quality spectra of quasars at redshift $z\sim2.5-2.9$ observed with the High Resolution Echelle Spectrometer on the Keck I telescope. We find that $\sim78\%$ of the systems identified by our algorithm are listed in the manual Voigt profile fitting catalogue. We demonstrate that the performance of our CNN is stable and consistent for all simulated and observed spectra with S/N $\gtrsim10$. Our model can therefore be consistently used to analyse the enormous number of both low and high S/N data available with current and future facilities.
Our CNN provides state-of-the-art predictions within the range $12.5\leq\log{N_{\mathrm{HI}}}/\mathrm{cm^{-2}}<15.5$ with a mean absolute error of $\Delta(\log{N}_{\mathrm{HI}}/{\rm cm}^{-2})=0.13$, $\Delta(z_{\mathrm{HI}})=2.7\times{10}^{-5}$, and $\Delta(b_{\mathrm{HI}})=4.1\ \mathrm{km\ s^{-1}}$.
The CNN prediction costs $<3$ minutes per model per spectrum with a size of 120\,000 pixels using a laptop computer. We demonstrate that CNNs can significantly increase the efficiency of analysing \Lya\ forest spectra, and thereby greatly increase the statistics of \Lya\ absorbers.
\end{abstract}


\section{Introduction}
\label{sec:intro}
The forest of neutral hydrogen (\HI) Lyman-$\alpha$ (\Lya) absorption lines imprinted on a quasar spectrum -- collectively known as the \Lya\ forest \citep{Lynds1971,Sargent1980} -- provides our best understanding of the intergalactic medium (IGM) and circumgalactic medium (CGM), on scales of tens to hundreds of kpc and to Mpc \citep{Cristiani1995,Fang1996}. The photons emitted by a background quasar are absorbed at the redshifted \Lya\ transition (rest-frame wavelength=1215.67\AA) in addition to higher order lines of the \HI\ Lyman series \citep[][]{Sargent1980}.

By number, \Lya\ absorption systems with low \HI\ column density dominate the \Lya\ forest and trace the underlying density of the \HI\ clouds \citep[e.g.][]{Schaye2001}. They can be used to probe the distribution and evolution of the baryonic matter, structure formation, and constrain cosmological parameters \citep[e.g.][also see reviews: \citealp{Rauch1998,Meiksin2009}]{Theuns1998,Theuns1999,Tytler2004,Lehner2007,Dave2010}. Additionally, the thermodynamic properties of these systems are primarily governed by two processes: (1) adiabatic cooling from the expansion of the Universe; and (2) photoheating by the ultraviolet background (UVB) light from quasars and galaxies \citep{Abel1999,Theuns2002,Bolton2009,Puchwein2015}. The competition between these two effects tracks the thermal state of the low-density IGM through a characteristic temperature-density relation \citep[e.g.][]{Hui1997,Haehnelt1998,Schaye1999,Schaye2000,Ricotti2000,Becker2007,Bolton2008,Rudie2012b}. Furthermore, the \Lya\ forest can also be used to probe cosmological models and constrain the properties of dark matter \citep[e.g.,][]{Viel2013,Baur2016,Garzilli2017,Irsic2017,Boera2019,Rogers2021}.

While the \Lya\ forest is easily identified in a quasar spectrum, the identification of \emph{individual} \Lya\ absorption systems within the forest is challenging. Conventionally, these absorption lines in the \Lya\ forest are fit with Voigt profiles\footnote{For example, the commonly used \textsc{vpfit} package, which is available from: \url{https://people.ast.cam.ac.uk/~rfc/vpfit.html}.} \citep[e.g.][]{Kim2002,Kim2013,Kim2021,Prochaska2005,Prochaska2009,Rudie2012a}; however, a manual fit to the entire \Lya\ forest is very time-consuming, and requires the aid of visual inspection, and many human hours. To avoid human bias, there are also studies that have developed automated Voigt profile fitting algorithms\footnote{We provide a few example codes here, but note that many efforts to generate an automated approach are unpublished. This problem is difficult, and an automated solution is not currently at the same level of accuracy that a human can produce.} \citep[][]{Romeel1997,Carswell2014,Bainbridge2017,Gaikwad2017}.

With future surveys and facilities such as the WHT Enhanced Area Velocity Explorer (WEAVE; \citealt{weaveqso}), and the 4-metre Multi-Object Spectroscopic Telescope (4MOST; \citealt{4MOST2019}), thousands of high resolution ($R\simeq20000$) quasar spectra are expected in the coming years. It will therefore not be feasible to analyse the enormous number of quasar spectra using conventional analysis methods. To overcome big data problems, such as this, machine learning techniques are essential. 

Machine learning techniques, in particular deep learning \citep{lecun2015deep}, have been widely applied to a variety of galaxy studies such as galaxy morphology \citep{Cheng2020a,Cheng2021b,Walmsely2022}, galaxy merger \citep{Bottrell2019,Ferreira2020}, and strong gravitational lensing \citep{Metcalf2019,Cheng2020b,Pearson2021}. Applications to analyse spectroscopic data or time-series data include gravitational wave analyses \citep{George2018}, transient objects \citep{Muthukrishna2019}, and spectral classification \citep{Bailer-Jones1998}. Recently, there has been a growing interest in applying machine learning techniques to the \Lya\ forest, including: (1) a \Lya\ forest emulator \citep{Bird2019,Rogers2019}; and (2) the identification and properties of damped \Lya\ systems \citep[DLAs;][]{Garnett2017,Parks2018,Wang2022}. DLAs are defined to have \HI\ column densities that exceed $N_{\mathrm{HI}}$ $\ge10^{20.3}\ \mathrm{cm^{-2}}$, and are easily identified by their strong, damped absorption features super-imposed on the \Lya\ forest. Unlike DLAs, the low \HI\ column density \Lya\ absorption systems associated with the \Lya\ forest ($N_{\mathrm{HI}}$ $<10^{17}\ \mathrm{cm^{-2}}$) have a relatively shallow depth and narrow absorption features. Furthermore, \Lya\ forest absorption features outnumber DLA absorption lines by orders of magnitude, and occupy a wider range of column density. These absorption lines are also often blended and confused with metal lines, making this a challenging and laborious problem. As a result, an efficient and reliable machine learning based solution to harvest the Lya forest -- both line detection and characterisation -- does not exist. Given the utility of these low column density \Lya\ systems in studying the physics of the IGM, it is essential to develop a machine-learning-based detection algorithm to identify and characterise these features in preparation for the coming `Big Data' era.

In this paper, for the first time, we apply a convolutional neural network (CNN) to efficiently identify \Lya\ forest systems ($N_{\mathrm{HI}}$ $<10^{17}\ \mathrm{cm^{-2}}$) and extract their physical properties, including the redshift, Doppler width, and \HI\ column density. While our primary goal is to efficiently extract the properties of the observed \Lya\ forest, our algorithm can also be used to identify \Lya\ absorbers in simulated spectra.
Since our approach is general, this allows a more direct comparison between spectra extracted from state-of-the-art hydrodynamic cosmological simulations and observations.
The paper is arranged as follows. Section~\ref{sec:qso_spectra} describes the generation of our simulated quasar spectra for training and initial testing purposes, and the observed quasar spectra that are used to validate our CNN predictions. Section~\ref{sec:DLmodel} explains the CNN models and the training strategies, and we describe the evaluation metric in Section~\ref{sec:evalution_metrics}. In Section~\ref{sec:result}, we test our pre-trained model with the simulated spectra, while in Section~\ref{sec:predict_2_obsspec}, we apply the CNN models to predict the parameters of the \Lya\ forest from observed spectra, and compare the CNN's predictions with the results based on Voigt profile fitting and human inspection from \citet[][hereafter R12]{Rudie2012a}. Finally, our conclusions are summarised in Section~\ref{sec:summary}.

\section{Quasar Spectra}
\label{sec:qso_spectra}
In this section we describe the simulated and observed quasar \Lya\ forest data that are used to train and test our network. While the technique that we employ can be readily applied to quasars at any redshift, the focus of our work is to study \Lya\ absorption in the optical wavelength range. Since the \Lya\ forest is blueward of the quasar \Lya\ emission line, to detect \Lya\ forest absorption features in the optical range (i.e. $\sim$3200\AA~to 7200\AA), the emission redshift of the quasars is in the range $z=1.6-5$. To satisfy the observed wavelength range, we generate simulated spectra at $z=3$ for training our CNN. The details of the spectrum generation are outlined in Section~\ref{sec:mock_spectra}, while in Section~\ref{sec:lya_sys_pixel}, we describe the pixel-level labelling of each \Lya\ absorption system. The observed spectra used to validate our model are described in Section~\ref{sec:observed_data}.

\subsection{Mock Spectra}
\label{sec:mock_spectra}
\begin{figure*}{}
\begin{center}
\graphicspath{}
	\includegraphics[width=2.1\columnwidth]{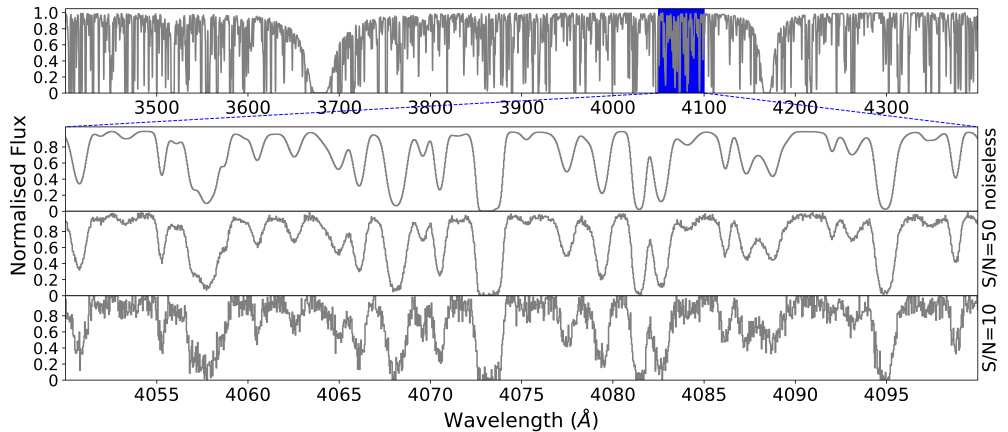}
   	\caption{The top panel shows a simulated \Lya\ forest spectrum of a quasar at redshift $z=3$ (no noise). The three subsequent panels show a zoom-in of the top panel (see blue box, top panel) with different signal-to-noise ratios (${\rm S/N} = \infty$, $50$, and $10$ per pixel for the second, third and fourth panels, respectively).}
    \label{fig:mock_spec}
\end{center}
\end{figure*}
The number of human-analysed quasar spectra that have been fit with Voigt profiles is currently limited by the time effort required to carefully analyse and fit each individual absorption line in every quasar spectrum. The quasar spectra that have been analysed are subject to human choices that may not reflect the true underlying properties of the absorption lines. For this reason, our training data are based on simulated quasar spectra to provide a large quantity of spectra together with ground-truth identifications of \Lya\ systems and their properties. Our simulated spectra were generated using packages in the {\sc pyigm} software\footnote{Primary Builders include: J. Xavier Prochaska, N. Tejos, and J. Burchett (\url{https://github.com/pyigm/pyigm}). We also implemented a minor change to this code; when generating Voigt profiles, we constructed a sub-pixellated wavelength array to sample each native pixel by ten sub-pixels. This accounts for the curvature of the profile within each pixel.}. The generated spectra represent a typical quasar at redshift $z=3$ and are convolved with an instrumental full-width at half-maximum (FWHM) resolution of $v_{\rm FWHM}=7\ \mathrm{km\ s^{-1}}$. These choices are motivated by the typical properties of high resolution spectra of quasars in current observatory archives. The velocity per pixel of these spectra is set to $2.5\ \mathrm{km\ s^{-1}\ pixel^{-1}}$. The impact of the model trained with this setup on predicting spectra with different assumptions for the properties of the spectra are discussed in Appendix~\ref{app:diff_spec}.

A catalog of \Lya\ forest absorption lines are drawn randomly from the column density distribution function (CDDF), $f\mathrm{(}N_{\mathrm{HI},X}\mathrm{)}$, following the default form implemented in \textsc{pyigm} (the Hermite spline model of \citealt{Prochaska2014}), where $X$ is the absorption distance. This provides a distribution of \HI\ absorption systems with $N_{\mathrm{HI}}$ $=10^{12}-10^{22}\ \mathrm{cm^{-2}}$ that can be imprinted onto a simulated quasar spectrum to generate absorption features with `ground-truth' labels (see Section~\ref{sec:lya_sys_pixel}). Note that this model was constrained at redshift $z\approx2.5$. \textsc{pyigm} uses inverse transform sampling of the $z=2.4$ column density distribution function to generate a list of \HI\ column densities; the corresponding Doppler parameters are drawn from the \citet{HuiRutledge1999} distribution. The redshifts of the mock lines are generated by inverse transform sampling the redshift-dependent incidence of absorption systems, $l(z)$. Finally, the spectra are generated without noise; additional noise is added later to test the sensitivity of our model to the adopted S/N (Section~\ref{sec:data_input}). In Fig.~\ref{fig:mock_spec}, we show an example of a simulated spectrum with different choices of the S/N. Our simulated spectra contain only the absorption lines of the \HI\ Lyman series, and do not include metal lines.

Machine learning applications commonly require training samples with a well-defined structure and clear corresponding labels, if possible. Since \Lya\ absorption features are relatively simple and have a well-defined structure that can be derived by only a few physical properties, i.e. Voigt profiles, having robust labels are more crucial than complexity of dataset to avoid confusion in a classification task. Hence, as a first attempt, this training dataset defines a clear structure of \Lya\ absorbers that helps a machine to draw a cleaner decision boundary in a high dimensional parameter space. Additionally, it helps us to analyse the performance of our automated algorithm and identify its limitations. As an alternative, we could generate simulated spectra with cosmological hydrodynamic simulations to account for the clustering and complex structure that exists in a real quasar spectrum. However, since the CGM structures in these simulations are unresolved \citep[][]{Rudie2019,Hummels2019,VanDeVoort2019}, it might be more sensible to train a machine using observed spectra in future works to account for the clustering of absorbers.

\subsection{\Lya\ absorption systems}
\label{sec:lya_sys_pixel}
\begin{figure}{}
\begin{center}
\graphicspath{}
	\includegraphics[width=\columnwidth]{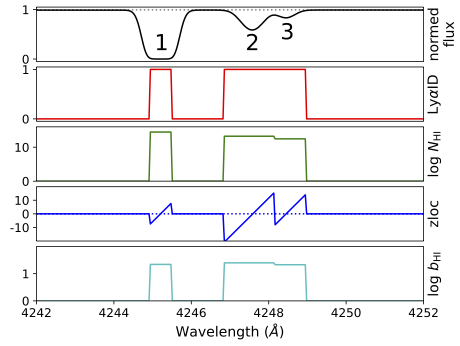}
   	\caption{Example of the training labels: \Lya ID, ${\log}N_{\mathrm{HI}}$, zloc, and ${\log}b_{\mathrm{HI}}$ in pixel scale from top to bottom. The grey dotted line in the top panel represents the normalised quasar continuum level, and the blue dotted line in the third panel shows zloc$=0$. Note that the training labels are only defined when \Lya ID=1. The absorption features are labeled 1, 2, and 3, ordered by column density.}
    \label{fig:label_in_pixel}
\end{center}
\end{figure}
The \Lya\ absorption features in a spectrum can be described with three physical properties: (1) the total \HI\ column density ($N_{\mathrm{HI}}$; $\mathrm{cm^{-2}}$), (2) the redshift ($z_{\mathrm{HI}}$) of the \HI\ absorbers, and (3) the Doppler width ($b_\mathrm{HI}$; $\mathrm{km\ s^{-1}}$). The $f\mathrm{(}N_{\mathrm{HI}}\mathrm{)}$ model provides a distribution of \HI\ absorbers that samples the \HI\ column densities of the \Lya\ forest. With the `ground-truth' information of the three aforementioned properties, we generated four labelling arrays for each pixel in the quasar spectrum (these labels are illustrated in Fig~\ref{fig:label_in_pixel}):
\begin{itemize}
    \item \textbf{Ly$\boldsymbol{\alpha}$ID}: set to a value of 1 if a \Lya\ absorber exists in this pixel, and 0 if not;
    \item $\boldsymbol{{\log}N_{\mathrm{HI}}}$: \HI\ column density (in units of $\mathrm{cm^{-2}}$) of the corresponding \Lya\ absorber on a logarithmic scale;
    \item \textbf{zloc}: the relative location of the centre of an absorption feature (in units of pixels\footnote{Note that zloc is a floating point number, since the centre of the associated absorption line is not coincident with the centre of a pixel.}). A pixel centred on an absorption feature is set to 0, and negative and positive values to pixels at the left and right, respectively. For example, if the centre of a given pixel is 2.4 pixels to the left of the centre of an absorption profile, we assign the label of this pixel to be $-2.4$;
    \item $\boldsymbol{{\log}b_{\mathrm{HI}}}$: Doppler width of the corresponding \Lya\ absorber on a logarithmic scale ($\mathrm{km\ s^{-1}}$).
\end{itemize}

\noindent First, to ensure that the absorption features used to train our machine are \Lya\ lines, we applied a cut to exclude the pixels with wavelengths where the \Lyb\ transition of the highest redshift \HI\ absorber appears. The initial pixel values for the four training label arrays were set to 0. The labels were generated for all \Lya\ systems ordered from the highest \HI\ column density to the lowest \HI\ column density. For each \Lya\ system, we first check if the optical depth, $\tau=N_{\rm HI}\,\sigma_{\alpha}$ ($\sigma_{\alpha}$ is the absorption cross-section for the \Lya\ transition), of the pixel is high enough to saturate the absorption line using a criterion of $\exp\mathrm{(}{-\tau}\mathrm{)}<0.015$, where the threshold is defined by $3/\mathrm{(S/N)}$ (where our fiducial S/N=200). If any pixel satisfies this criterion, we store the \Lya ID, ${\log}N_{\mathrm{HI}}$, zloc, and ${\log}b_{\mathrm{HI}}$ of this absorber in the label arrays. Note that `zloc' represents the location of the centre of an absorption feature, where the centre (zloc$=0$) is drawn using the redshift of the \Lya\ system. If the listed \Lya\ system does not saturate a pixel, 
it is then used to provide values to the corresponding pixels where the flux of the absorption features is $<0.995$. Note, if multiple absorption components contribute to the total optical depth in a pixel, we labelled only the dominant line. This means that in this work we do not consider the impact of a secondary or additional line blends in a single pixel. A more thorough investigation about the effect of blended lines will be carried out in future work. Fig.~\ref{fig:label_in_pixel} shows an example of the labelling procedure that we use in this work. In the example shown in Fig.~\ref{fig:label_in_pixel}, labels are first assigned to the leftmost (strongest) feature, i.e. feature 1. Every pixel associated with this absorption line that has a flux less than $0.015$ is assigned a \Lya ID$=1$; the column density and Doppler parameter is the same for all of the associated pixels of this feature, and the zloc label represents the non-integer pixel difference from the centre of the absorption line profile. The next strongest absorption line, feature 2, is then labelled; because the central optical depth is not saturated we label all pixels that have a flux $<0.995$. The rightmost feature 3, which is partially blended, is labelled using the same approach, however, the labels are only applied to the pixels where the pixel optical depth contributed by this feature is highest.

\subsection{Archival Quasar Observations}
\label{sec:observed_data}
To validate our machine's prediction on real data, we use the 15 quasar spectra observed and reduced by R12. These data were observed with the High Resolution Echelle Spectrometer \citep[HIRES;][]{Vogt1994} on the Keck I telescope. The redshifts of these quasars are in the range $2.5\lesssim z\lesssim2.9$, and the spectra have $R\ \cong\ 45\,000$ ($v_{\rm FWHM}\ \cong\ 7\ \mathrm{km\ s^{-1}}$), high signal-to-noise ratio (${\rm S/N}\sim\ 50-200\ \mathrm{pixel^{-1}}$), and cover the wavelength range 3100$-$6000\AA. We resampled these spectra to $2.5\ \mathrm{km\ s^{-1}\ pixel^{-1}}$ (while conserving flux) to be consistent with the input of our CNN model (see Section~\ref{sec:data_input} and Appendix~\ref{app:diff_spec}). Further details about the observations and data reduction procedure are outlined by R12\footnote{Some spectra contain DLA absorption lines. Our CNN model is sensitive to \Lya\ systems with low column density, and it then ignores the DLA features. Hence, these features do not impact the results.}. 

\section{Deep Learning Model}
\label{sec:DLmodel}
\begin{figure*}{}
\begin{center}
\graphicspath{}
	\includegraphics[width=2.1\columnwidth]{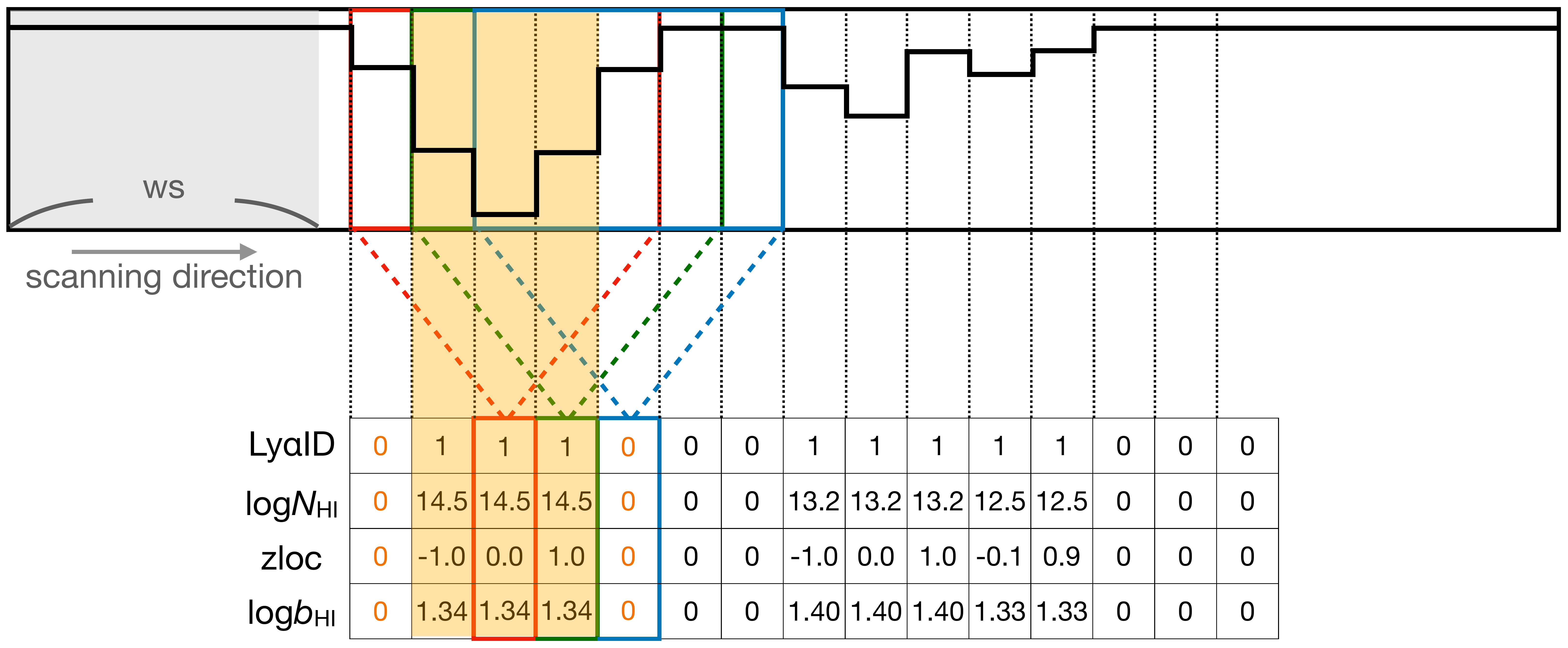}
   	\caption{Schematic diagram of the scanning process that we use to train the data. The scanning window size is $ws$, and the step size is 1 pixel. All of the pixels in a given window are used as input to the CNN, and the corresponding output label of each window is assigned to the pixel located at the centre of the window. The red, green, and blue boxes demonstrate how the corresponding labels change when the window shifts by 1 pixel. Labels are only assigned if the centre pixel of the window is within $cnpix=\pm1$ pixels of the centre of an absorption feature (this is represented by the yellow box). Outside of this defined area, the corresponding labels are set to 0. Note that the value of $cnpix$ is a hyperparameter of the network.}
    \label{fig:scanning}
\end{center}
\end{figure*}
We employ multi-task learning \citep{Caruana1998,Ruder2017} by training with and predicting four outputs (labels): \Lya ID, ${\log}N_{\mathrm{HI}}$, zloc, and ${\log}b_{\mathrm{HI}}$ (see Section~\ref{sec:lya_sys_pixel}). The network is generalised to approach these four tasks at the same time. The details of the CNN structure for our multi-task learning are described in Section~\ref{sec:cnn_design}. The prediction of each variable complements the prediction of the other variables by combining their losses  (details in Section~\ref{sec:loss_func})\footnote{The loss quantifies the difference between the expected output (i.e. truth) and the predicted output by a machine learning model, while the loss function is the function used to calculate the loss.} as part of the training process.

We employ similar training strategies to that adopted by \citet{Parks2018} to `scan' through a spectrum with a fixed-size window ($ws$) and a 1 pixel step size. To do this, we used the \texttt{fit\_generator} function in {\sc keras}. This method increases the machine's performance by analysing hundreds of pixels in a segmentation per step rather than tens of thousands of pixels in a whole spectrum in one go. A \texttt{fit\_generator} has the added benefit that each window is generated at run-time from the full spectrum, and therefore reduces the amount of VRAM required (or, equivalently, allows us to include more training data). The schematic diagram of the scanning process is shown in Fig.~\ref{fig:scanning}.

\subsection{Data Input}
\label{sec:data_input}
In each spectral window used as input (of size $ws$), there are four training labels, and these labels correspond to the properties of the centre pixel in this window. Our CNN is therefore trained with and only predicts the corresponding values at the central pixel within this window from each labelling array. For example, in Fig.~\ref{fig:scanning}, the labels that correspond to the red spectral window are listed in the red labels box, and the ones that correspond to the green spectral window are in the green labels box, etc. The size of the spectral window, $ws$, is a hyperparameter that is objectively selected using an optimisation algorithm (Section~\ref{sec:cnn_design}). We scan each training spectrum from left to right during each epoch. Each batch contains one spectral window from each training spectrum. This approach ensures that all training spectra are fully `scanned' and their training losses are taken into account in each epoch (see also Section~\ref{sec:loss_func}).

To ensure that the CNN prediction is primarily sensitive to absorption features that are located at the centre of the window, we define an additional hyperparameter, $cnpix$. This hyperparameter is defined by the absolute value of zloc, |zloc|$\leq cnpix$, and determines the pixels that are recognised as the `centre' of an absorption feature. For example, in Fig.~\ref{fig:scanning}, if $cnpix=1$, the yellow shaded area is defined as the `centre' region, and the true values outside this range are set to 0 as highlighted by the yellow labels. The CNN is trained with, and predicts the labels associated with, the central pixel of the window. The variable $cnpix$ ensures that the training process only learns from an absorption feature that overlaps with the pixel in the centre of a window.

Additionally, we noticed that training our machine with noiseless spectra results in a significantly worse performance when predicting a noisy spectrum (see Appendix~\ref{app:diff_snr}). To overcome this issue so that our machine can sensibly be applied to predict accurate labels to real data, we included additional noise to each spectrum. The S/N of a given spectrum is drawn from a Gaussian distribution, with a mean of 10 and a standard deviation of 2.  Given this S/N value, we randomly perturb every pixel in the perfect normalised spectrum by a Gaussian distribution with a standard deviation of 1/(S/N). In previous studies, \Lya\ forest analyses have primarily relied on spectra with S/N $>20$. Hence, we chose a low S/N value as a typical value to allow our machine to produce reliable results when analysing observed quasar spectra that are of somewhat lower S/N. Appendix~\ref{app:diff_snr} outlines the tests we performed to validate this approach, and demonstrate that this stabilises the predictions for spectra with different S/N.


\subsection{CNN Architecture}
\label{sec:cnn_design}
\begin{figure*}{}
\begin{center}
\graphicspath{}
	\includegraphics[width=2.1\columnwidth]{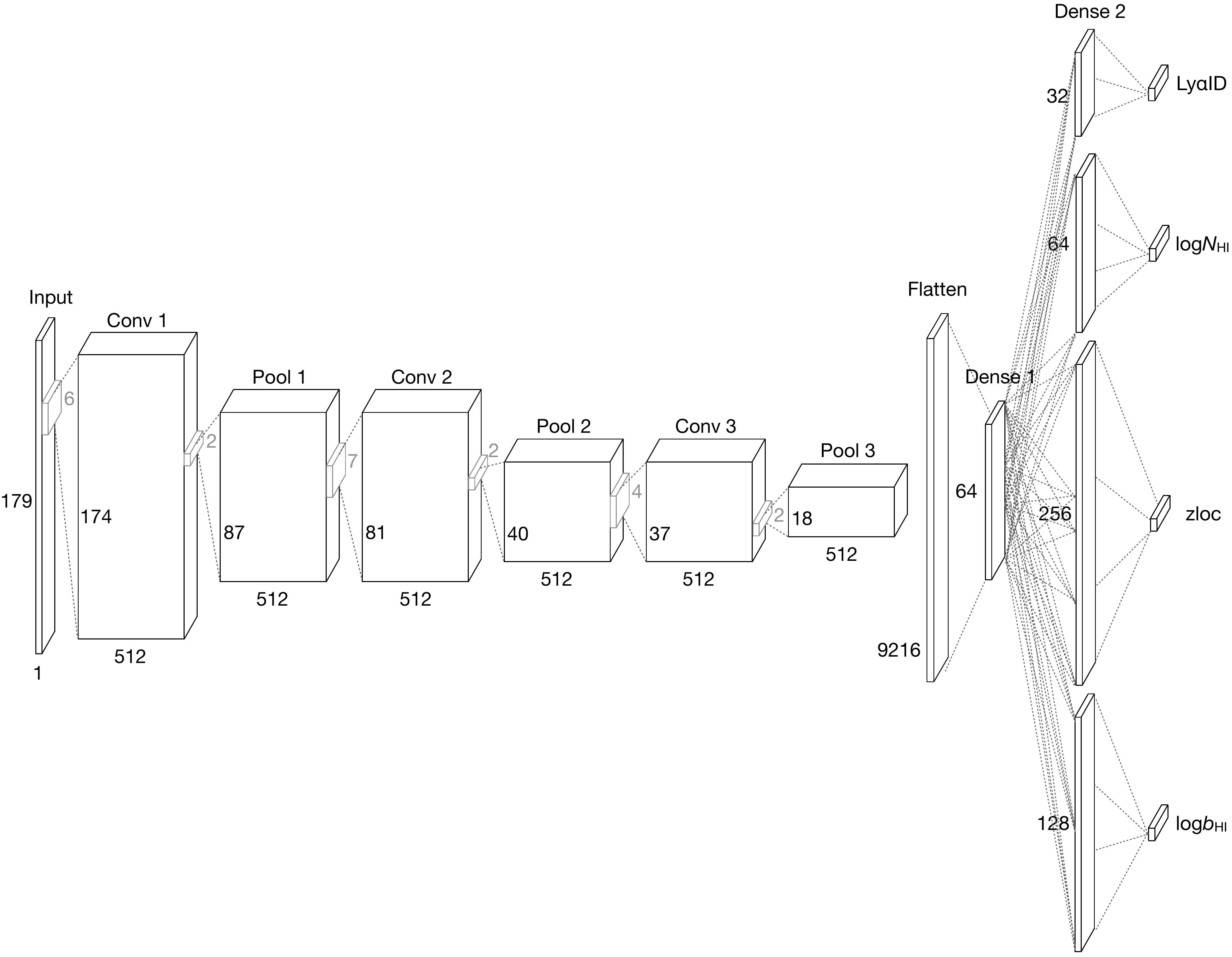}
   	\caption{Schematic diagram of the CNN architecture used in this work. It is composed of three 1-dimensional convolutional layers with pooling layers following each, one dense layer to connect each component, and four dense layers for four target outputs. The values of relevant hyperparameters are listed in Table~\ref{tab:hyper-parameters}.}
    \label{fig:cnn_design}
\end{center}
\end{figure*}
\begin{table}
	\centering
	\begin{tabular}{llc}
		\hline
		\multicolumn{1}{c}{} & {Hyperparameters} & {Optimised value}\\
		\hline\hline
		\multicolumn{1}{l}{Data Input} & {window size ($ws$)} & {179} \\
		\multicolumn{1}{l}{} & {central pixels ($cnpix$)} & {1} \\
		\hline
		\multicolumn{1}{l}{CNN} & {L2} & {0.0} \\
		\multicolumn{1}{l}{Architecture} & {dropout} & {0.1} \\
		\multicolumn{1}{l}{} & {conv\_filter\_1} & {512} \\
		\multicolumn{1}{l}{} & {conv\_filter\_2} & {512} \\
		\multicolumn{1}{l}{} & {conv\_filter\_3} & {512} \\
		\multicolumn{1}{l}{} & {conv\_kernel\_1} & {6} \\
		\multicolumn{1}{l}{} & {conv\_kernel\_2} & {7} \\
		\multicolumn{1}{l}{} & {conv\_kernel\_3} & {4} \\
		\multicolumn{1}{l}{} & {dense\_1} & {64} \\
		\multicolumn{1}{l}{} & {dense\_2\_ID} & {32} \\
		\multicolumn{1}{l}{} & {dense\_2\_N} & {64} \\
        \multicolumn{1}{l}{} & {dense\_2\_z} & {256} \\
		\multicolumn{1}{l}{} & {dense\_2\_b} & {128} \\
		\hline
	\end{tabular}
	\caption{Hyperparameters used in our CNN architecture. These values are selected using a Bayesian optimisation algorithm \citep{Snoek2012}.}
	\label{tab:hyper-parameters}
\end{table}
Fig.~\ref{fig:cnn_design} shows our CNN architecture, which follows the same form as the one used in \citet{Parks2018}, including three 1-dimensional convolutional layers (i.e. Conv 1, Conv 2, Conv 3) and each of them is followed by a pooling layer with a kernal size of 2. A dropout is inserted after the third pooling layer (Pool 3), and the array is flattened to connect with a dense layer (Dense 1). Four separate dense layers are then connected with the `Dense 1' layer and dropouts are applied to each dense layer. The dropout rate is consistent throughout the network and is one of the hyperparameters that is selected with an optimisation algorithm.

The activation function used before the output layer is consistently \texttt{ReLu}: $f\left( x \right)=max\left( 0,x \right)$ \citep{Agarap2018}, and the activation functions for the outputs depend on the desired output range of the target variables. Hence, for \Lya ID, we applied the \texttt{sigmoid} function: $f\left( x \right)=1/(1+e^{-x})$, which outputs a value between 0 and 1 as a probability. For zloc we adopted a \texttt{linear} function: $f\left( x \right)=x$, and for both ${\log}b_{\mathrm{HI}}$\footnote{The minimal value of $b_{\mathrm{HI}}$ in this work is $15\ \mathrm{km\ s^{-1}}$. Hence, the logarithmic value is always $>0$.} and ${\log}N_{\mathrm{HI}}$ we used \texttt{ReLu}, which outputs a value $f\left( x \right)={\rm max}(0,x)$. Several crucial hyperparameters in our CNN architecture were objectively selected by a Bayesian optimisation process \citep[][also see appendix~\ref{app:bayesian_opt}]{Snoek2012} over a range of possible values. The results are listed in Table~\ref{tab:hyper-parameters}. In addition to the hyperparameters of the CNN architecture, we include two additional hyperparameters from Section~\ref{sec:data_input}: (1) the window size ($ws$) and (2) the number of pixels that are used to define the centre of an absorption feature ($cnpix$). These two hyperparameters are critical in determining the types of \Lya\ systems that our CNN is sensitive to\footnote{One can use a larger size of scanning window to help improve the sensitivity in detecting systems with higher column density. Note that these systems are fewer. To carry out this optimisation, one also needs to consider the issues of strongly imbalanced number of different systems.}; the values of these parameters depend on the science question being addressed. We therefore use a Bayesian optimisation process to decide their values without human intervention.

Finally, the learning rate was set to 0.0001 and we applied the {\texttt{Adam}} optimiser \citep{Kingma2015}. The maximal number of iteration for each training is 20 epochs, but only the model with the minimal validation loss within the 20 epochs is saved.

\subsection{Loss Function}
\label{sec:loss_func}
With our multi-task learning model, four outputs were produced for a given input: \Lya ID, ${\log}N_{\mathrm{HI}}$, zloc, ${\log}b_{\mathrm{HI}}$. With the \texttt{fit\_generator} function, the final loss per epoch for each output is an average value of the losses of all steps. For a binary classification task, the loss of `\Lya ID' uses a binary cross-entropy loss function:
\begin{equation}
    L_{ID}=-\frac{1}{N}\sum_{i=1}^{N}y_{c,i}\log\left( p_{c,i} \right)+\left( 1-y_{c,i} \right)\log\left( 1-p_{c,i} \right),
\end{equation}
\noindent where $N$ is the total number of training windows per epoch, i.e. the number of input training spectra ($=$ number of data per step) times the number of pixels in each spectrum ($=$ number of steps)\footnote{Recall that the model is trained by scanning through all spectra.}, $y_c$ represents the true classification label (i.e. \Lya ID$=y_c=1$ for a \Lya\ absorption system), and $p_c$ is the probability of being a \Lya\ system predicted by the CNN.
The loss functions of the remaining outputs use a masked mean square error (MSE):
\begin{equation}
    L_{j}=\frac{1}{N^{'}}\sum_{i=1}^{N}y_{c,i}\left( y_{j,i}-\hat{y}_{j,i} \right)^{2},
\end{equation}
\noindent where $N^{'}$ is the total number of training windows per epoch where $y_c$ equals to 1, and $j=\left\{ N_{\mathrm{HI}},z,b_{\mathrm{HI}} \right\}$ represents the loss functions of ${\log}N_{\mathrm{HI}}$, zloc, ${\log}b_{\mathrm{HI}}$, respectively. The $y_{j,i}$ are the true values of $j=\left\{ N_{\rm HI},z,b_{\rm HI} \right\}$, while the $\hat{y}_{j,i}$ are the predicted values from the CNN. With this `masked' form of the loss function for $\log N_{\rm HI}$, $z$, and $\log b_{\rm HI}$, losses are only contributed to the final loss per epoch when $y_{c}=1$.\footnote{We note that this masked loss function ensures that our machine is not biased by the $\log N_{\rm HI}$, $z$, and $\log b_{\rm HI}$ labels in pixels where there is no absorption.} The final loss function of the CNN training process per epoch is the sum of the above-mentioned losses:
\begin{equation}
L=L_{ID}+L_{N_{\rm HI}}+L_{z}+L_{b_{\rm HI}}
\end{equation}
\noindent Note that the scale of each loss needs to be comparable in order to prevent a biased weighting due to a single label that contributes most of the loss. For example, in our preliminary test, we found that a large uncertainty in predicting linear $b_{\mathrm{HI}}$ values (range of $15-75$ $\mathrm{km\ s^{-1}}$) contributes a significant loss which therefore decreases the CNN's capability of precising predicting the other labels. Hence, we opted to predict the logarithmic $b_{\mathrm{HI}}$ values in this work.

\section{Evaluation Metrics}
\label{sec:evalution_metrics}
Before showing the results of our CNN models, we first introduce the metrics that were used to evaluate the CNN performance. For the classification of \Lya\ absorbers, we use recall and precision, as defined below, to evaluate the CNN performance.
\begin{equation}
\label{eq:prec_rec}
    	\mathrm{recall}=\frac { \mathrm{TP} }{ \mathrm{TP}+\mathrm{FN} } ;\quad \mathrm{precision}=\frac { \mathrm{TP} }{ \mathrm{TP}+\mathrm{FP} },
\end{equation}
\noindent where `TP' is a true positive (i.e. a correct classification), `FP' is a false positive (i.e. a mis-classified system), and `FN' is false negative corresponding to true systems that are missed by our CNN. Recall is a measure of completeness: the fraction of true absorbers identified by the CNN. Precision is a measure of the fraction of identified systems that are real. We have designed the CNN to have high precision at the expense of recall, so that we are confident that a CNN-classified \Lya\ system is a real \Lya\ system. This choice may need to be different, depending on the scientific question being addressed.


On the other hand, when estimating the physical properties of a \Lya\ absorber such as redshift, \HI\ column density, and Doppler width, we consider two metrics: (1) the root mean square error (RMSE) and (2) mean absolute error (MAE), to assess the `accuracy' of the CNN predictions. The RMSE is defined as:
\begin{equation}
\label{eq:RMSE}
    RMSE=\sqrt{\frac{\sum_{k=1}^{N_{\mathrm{Ly}\alpha}}\left( y_{k}-\hat{y}_{k} \right)^{2}}{N_{\mathrm{Ly}\alpha}}},
\end{equation}
\noindent where $N_{\mathrm{Ly}\alpha}$ is the number of matched \Lya\ systems, and $y_k$ and $\hat{y}_k$ represent the `true' and `predicted' values of each \Lya\ system, respectively. The RMSE is strongly impacted by the outliers due to the square of the residual. Hence, we also introduce MAE (Equation~\ref{eq:MAE}) which is more resilient to outliers than the RMSE.
\begin{equation}
\label{eq:MAE}
    MAE=\frac{\sum_{k=1}^{N_{Ly\alpha}}\left| y_{k}-\hat{y}_{k} \right|}{N_{Ly\alpha}},
\end{equation}
\noindent where the definition of each variable is the same as Equation~\ref{eq:RMSE}. The MAE is more useful, since we do not expect the CNN to be absolutely correct. For example, in many cases our CNN predicts that a single \Lya\ absorber is required to recover an absorption feature, while there are in fact many neighbouring lines that contribute to this absorption feature (see discussion in Section~\ref{sec:mock_FN}). For this example, we will have poor estimates of the physical properties when comparing with the `true' values, and this yields strong outliers. Thus, the MAE is a more robust indicator of the CNN performance than the RMSE in the context of this study. In later sections, we will list both quantities, but the discussion will be based on the MAE.

\section{Prediction To Simulated Spectra}
\label{sec:result}
\begin{figure*}{}
\begin{center}
\graphicspath{}
	\includegraphics[width=1.9\columnwidth]{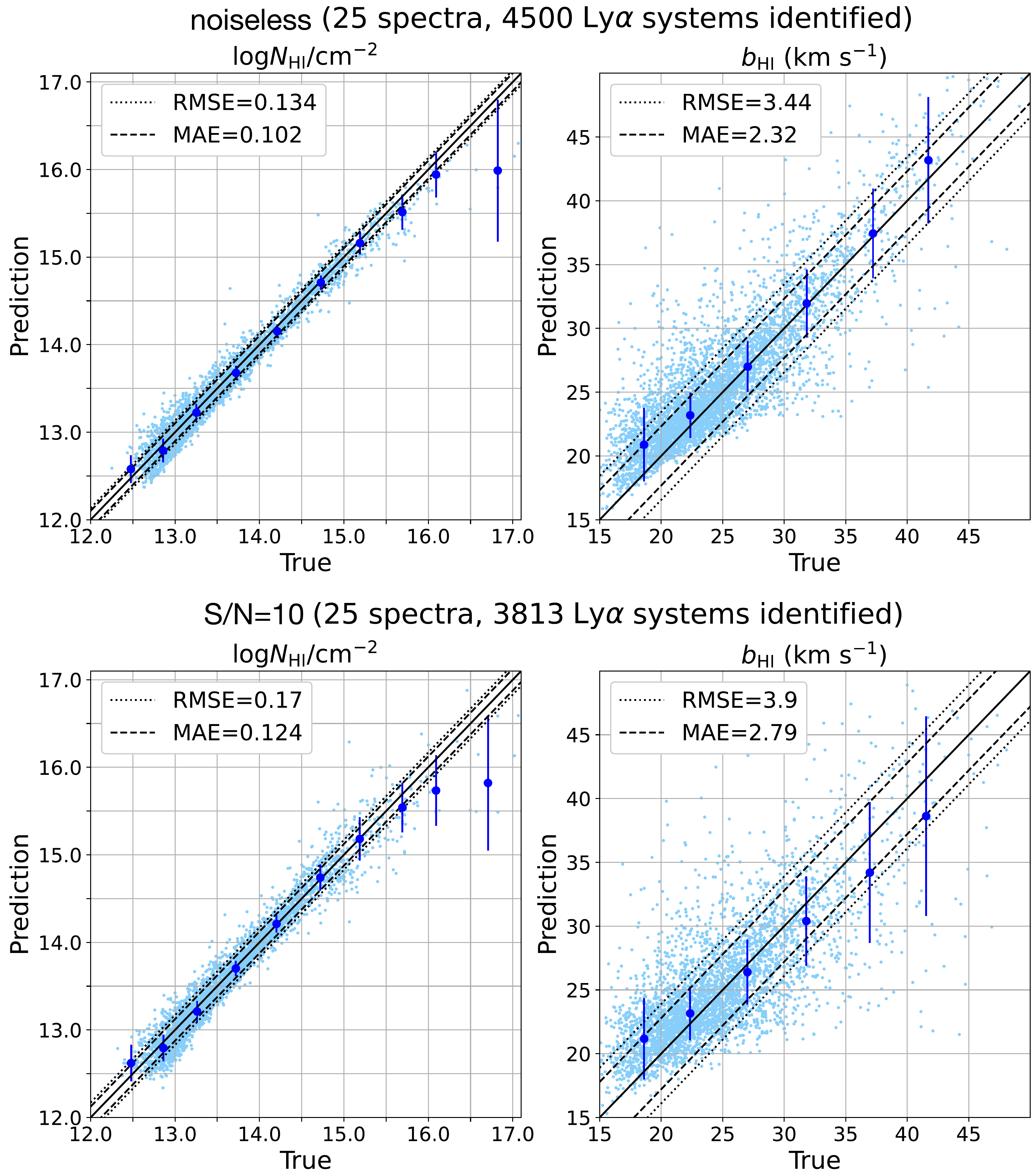}
   	\caption{Comparisons between the true and predicted values of $\log{N}_{\mathrm{HI}}/{\rm cm}^{-2}$ and $b_{\mathrm{HI}}$ with \Lya ID $>0.3$ using noiseless spectra (top) and spectra with S/N $=10$ (bottom). The black solid line shows $f\left( x \right)=x$, the dashed lines indicate the scatter range defined by the MAE of each plot, and the dotted lines are the range defined by the RMSE of each plot. Dark blue datapoints show the median values within different bins of the true values. The $y$-axis error bar presents the MAE of each bin.}
    \label{fig:mock_N_b_1to1}
\end{center}
\end{figure*}

With the aforementioned setups in Section~\ref{sec:DLmodel}, we trained a CNN model with 900 simulated spectra\footnote{This number of training set is sufficient since each spectrum includes over 20\,000 segmentation windows for the training process. Using additional spectra did not improve our result.} with a S/N randomly drawn from a Gaussian distribution with a mean of 10 and a standard deviation of 2. An independent set of 25 simulated spectra (a noiseless set and a S/N $=10$ set) were used to examine our pre-trained CNN model.

\subsection{CNN-classified \Lya\ forest systems}
\label{sec:cnn_lya_mockspec}
\begin{table*}
	\centering
	\begin{tabular}{lccccc}
		\hline
		\multicolumn{1}{l}{Test Sets} & {Precision} & {Recall} & {$\Delta{\log{N_{\mathrm{HI}}}/{\rm cm}^{-2}}$} & {$\Delta{z_{\mathrm{HI}}}$} & {$\Delta{b_{\mathrm{HI}}}$ ($\mathrm{km\ s^{-1}}$)}\\
		\hline\hline
		\multicolumn{1}{l}{noiseless mock spectra (\Lya ID $>0.7$)} & {0.992} & {0.127} & {0.08} & {1.7$\times10^{-5}$} & {1.4} \\
		\multicolumn{1}{l}{noiseless mock spectra (\Lya ID $>0.3$)} & {0.994} & {0.322} & {0.10} & {2.4$\times10^{-5}$} & {2.3} \\
        \hline
  		\multicolumn{1}{l}{S/N $=10$ mock spectra (\Lya ID $>0.3$)} & {0.987} & {0.273} & {0.12} & {3.0$\times10^{-5}$} & {2.8} \\
        \hline
		\multicolumn{1}{l}{\bf{R12 spectra (\Lya ID $>0.3$):}} & {} & {} & {} & {} & {} \\
        \multicolumn{1}{l}{All predictions} & {0.782} & {0.260} & {0.14} & {2.7$\times10^{-5}$} & {4.2} \\
        \multicolumn{1}{l}{$12.5\leq\log{N_{\mathrm{HI}}}/\mathrm{cm^{-2}}<15.5$} & {0.792} & {0.258} & {0.13} & {2.7$\times10^{-5}$} & {4.1} \\
		\hline
	\end{tabular}
	\caption{Evaluation metrics (precision, recall, MAE) of CNN-classified systems on different test datasets. All of the results tabulated here are based on the same model, which is trained on noisy spectra (where the S/N is drawn from a Gaussian distribution with a mean $=10$ and a standard deviation $=2$).}
	\label{tab:evaluation_diffsets}
\end{table*}
With a CNN prediction for each pixel in all spectra, we used the following two criteria to identify \Lya\ systems: (1) \Lya ID $>0.3$, and (2) |zloc|$\leq{cnpix}$, where $cnpix=1$. The former criterion judges if a pixel contains a \Lya\ system by the binary classification probability. The initial probability threshold $>0.3$ used for \Lya ID is considered to have the maximum number of identified systems without decreasing the precision by selecting pixels with low predicted probabilities. The second criterion is applied in order to identify the centre of an absorber.


To compare the CNN-classified \Lya\ systems with the ground-truth labels, we match the input and predicted catalogue of systems; our matching criteria require that the velocity difference between the input and prediction is smaller than half of the minimum FWHM that can be detected by a machine. This FWHM threshold is estimated by the minimum $b_{\mathrm{HI}}$ value our CNN predictor can detect, i.e., $b_{\rm min}=15\ \mathrm{km\ s^{-1}}$, using the relation: $\mathrm{FWHM}=2\sqrt{\log(2)}\,b$. Hence, the threshold applied is $\sqrt{\log(2)}\,b_{\rm min}\sim12.5\ \mathrm{km\ s^{-1}}$.

The comparisons between the true and predicted systems with \Lya ID $>0.3$ using noiseless spectra and S/N $=10$ spectra are shown in Fig.~\ref{fig:mock_N_b_1to1}. This figure provides an indication of the upper and lower ranges of CNN predictions for mock spectra of different noise levels. With low S/N, the total number of matched systems decreases when using the same probability threshold. However, the overall CNN performance remains consistent, with only a minor increase of the MAE. We summarise the evaluation metrics of different datasets in Table~\ref{tab:evaluation_diffsets}.

When applying a higher probability threshold to \Lya ID for spectra of the same noise, fewer systems with a high accuracy are matched. For example, in Table~\ref{tab:evaluation_diffsets}, when applying \Lya ID $>0.7$ to predict noiseless spectra, the recall drops while our CNN predictions show an improvement.

For either noiseless or S/N $=10$ simulated spectra, the overall precision of our CNN is over 0.98. In the following sections, we investigate the causes of the FP and FN classifications.

\subsection{False Positive}
\label{sec:mock_FP}
\begin{figure}{}
\begin{center}
\graphicspath{}
	\includegraphics[width=\columnwidth]{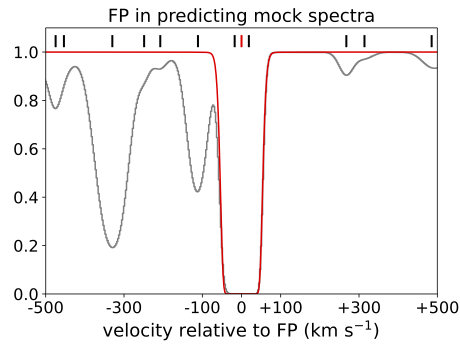}
   	\caption{An example of a false positive (red curve). The gray curve shows the contribution of all absorption lines. Short tick marks above the spectra indicate the centre of a \Lya\ system (black for catalogue lines, red for false positive). The velocity between the red prediction and the closest neighbour system from the true catalogue of primary absorption lines is $\Delta{V}=17.5~{\rm km~s}^{-1}$.}
    \label{fig:mock_FP_example}
\end{center}
\end{figure}
A false positive (FP) is an absorption system identified by our CNN classifier that cannot be matched to a \Lya\ system in the simulated true label catalogue. When predicting the labels of a simulated spectrum that only contains \Lya\ absorbers, our CNN reaches a precision of over 0.98 for spectra with S/N $=10$ (over 0.99 for noiseless spectra). The FP in this case is exclusively a simple mismatch due to the velocity threshold ($\sim12.5\ \mathrm{km\ s^{-1}}$) used in matching systems between true and predicted catalogues. An example is shown in Fig.~\ref{fig:mock_FP_example}. The velocity difference between our CNN-classified system and the closest neighbour is $17.5\ \mathrm{km\ s^{-1}}$ in this example. FPs occur when an absorption feature comprises multiple nearby lines, while our CNN tends to use one line to describe the absorption feature. This results in a shift of the defined centre and the mismatch of the true and predicted \Lya\ systems. Through visual inspection, we noticed that the parameters of the FPs predicted by the CNN classifier fit the absorption feature as well as the true labels, especially given that our classifier is trained on data of S/N $\simeq10$. This type of failure can also happen when using conventional methods such as Voigt profile fitting (e.g. human bias, or indistinguishable absorption profiles). This reflects a potential underestimation of the number of \Lya\ systems that are indifferentiable due to confusion or insufficient S/N, or because they are due to sub-structures of \HI\ gas within a larger \HI\ gas cloud. This may be improved by including higher order Lyman series lines as part of the CNN training process in future work.

\subsection{False Negative}
\label{sec:mock_FN}
\begin{figure}{}
\begin{center}
\graphicspath{}
	\includegraphics[width=\columnwidth]{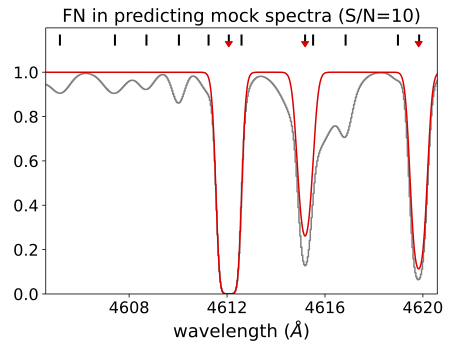}
	\includegraphics[width=\columnwidth]{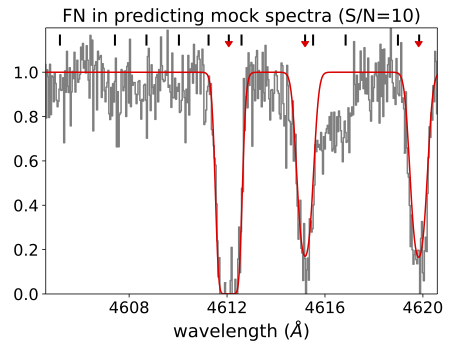}
   	\caption{Several examples of false negative. The centres of each true \Lya\ system are labelled by black short tick marks above the spectra, while the red arrow indicates a match between the true and predicted systems. Hence, a black tick mark without a matching red arrow represents a false negative. The top panel shows a noiseless spectrum and the bottom panel is a spectrum with S/N $=10$. The gray histogram represents the data and the red curve is a reconstruction based on all of the predicted lines by our CNN.}
    \label{fig:mock_FN_example}
\end{center}
\end{figure}
\begin{figure*}{}
\begin{center}
\graphicspath{}
	\includegraphics[width=2.1\columnwidth]{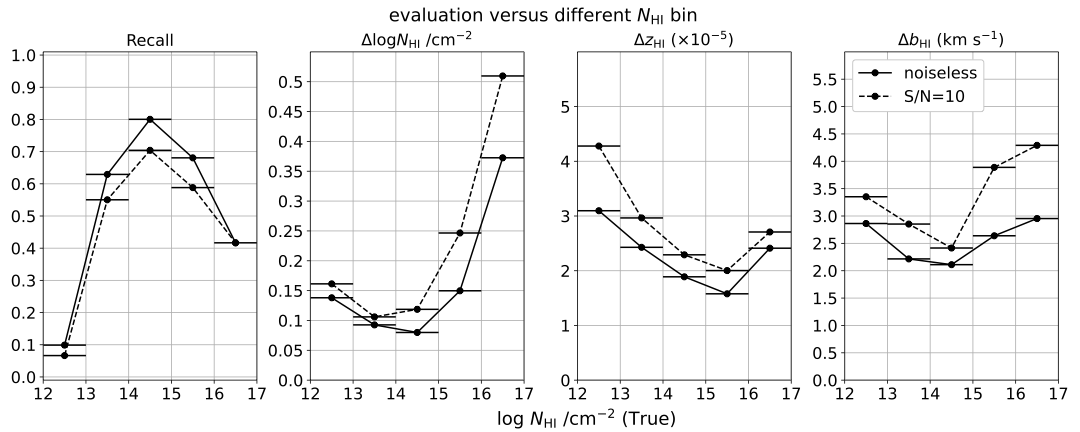}
  	\caption{Evaluation graphs of recall and the MAE of $\log{N_{\mathrm{HI}}}/{\rm cm}^{-2}$, redshift ($z_{\mathrm{HI}}$), and $b_{\mathrm{HI}}$ values, within different $\log{N_{\mathrm{HI}}}/{\rm cm}^{-2}$ bins. The horizontal lines of each data point represent the range of each bin.}
    \label{fig:eval_diffNHI}
\end{center}
\end{figure*}
False negatives (FN) occur when a system is listed in the true label catalogue, but it is not identified by our CNN. Examples are shown in Fig.~\ref{fig:mock_FN_example}. Since we train our CNN with noisy spectra, some detailed structures are buried in the noise, resulting in either a non-detection or low predicted probability (\Lya ID) to these pixels. This is a compromise between the accuracy and the feasibility of a CNN technique to spectra with low S/N. One can train a CNN with a higher S/N to increase the accuracy of a CNN detection and therefore reduce potential FN that are impacted by noise. In our test, when training our CNN with noiseless spectra\footnote{Note that this result used a CNN architecture with hyperparameters that were specifically tuned to noiseless data.}, the machine reaches a high recall ($\sim0.88$) and a high precision ($\sim0.90$) when testing on a noiseless spectrum. However, the ability to predict physical properties such as $b_{\mathrm{HI}}$ and $N_\mathrm{HI}$ drops significantly for a CNN that is trained on noiseless data, but applied to noisy spectra (for further details, see Appendix~\ref{app:diff_snr}). This severely limits the utility of a CNN model, since most spectroscopic data are of low S/N.

We summarise four main cases that contribute to false negatives:
\begin{enumerate}
    \item Weak absorption features that are not significantly detected in the noisy data (S/N $\sim10$) used for the training process. Our CNN then has difficulty distinguishing these features from the noise, and provides either a non-detection or a low predicted probability, i.e. \Lya ID $<0.3$, even for data with much higher S/N.
    \item A strong absorption feature composed of multiple neighbouring lines (e.g. see Fig.~\ref{fig:mock_FP_example}). This type of FN occurs when  our CNN uses one line to fit an absorption feature while this feature is in fact composed of several \Lya\ systems (Section~\ref{sec:mock_FP}). This mismatch therefore contributes several false negatives, and one false positive.
    \item A strong, broad absorption feature with a size that is larger than the scanning window, i.e. $\Delta V >447.5\ \mathrm{km\ s^{-1}}$. Due to the fixed size of our scanning window, our CNN is restricted to features that are well-defined within the window size.
    \item Complex absorption features contributed by many nearby lines. Similar to case (ii) above, our CNN only fits a dominant feature from this complex structure and misses other overlapped absorption features formed by nearby, usually weaker, \Lya\ systems.
\end{enumerate}
\noindent We find that the dominant FN contribution comes from weak absorption features that are within $3\sigma$ of the continuum; our 25 test spectra indicate this type of FN contributes $\sim86\%$ of the total number of FNs. These weak absorption features have an average value of ${\log}N_{\mathrm{HI}}/{\rm cm}^{-2}=12.40\pm0.25$ and $b=28.9\pm9.8~{\rm km~s}^{-1}$. If we exclude this kind of FN, the recall improves from $\sim0.32$ to $\sim0.77$ for noiseless spectra.

In Fig.~\ref{fig:eval_diffNHI} we present the change of recall and the MAE values of $\log{N_{\mathrm{HI}}}$, $z_{\mathrm{HI}}$, and $b_{\mathrm{HI}}$ grouped by column density using noiseless spectra (solid line) and S/N $=10$ spectra (dashed line). This demonstrates that our CNN model has better recall to \Lya\ systems with a column density range of $13\leq\log{N_{\mathrm{HI}}}/{\rm cm}^{-2}<16$. By visual inspection, we found that the false negatives for systems with column density in this range are only the cases (ii)$-$(iv) listed above. Compared to other bins, low \HI\ column density systems (i.e. $\log{N_{\mathrm{HI}}}/{\rm cm}^{-2}<13$) contribute weak absorption features which can be hidden in the noise, and result in much lower recall value ($\lesssim0.1$), i.e. a higher fraction of FNs.

\section{Application to observational data}
\label{sec:predict_2_obsspec}
Following the setup described in Section~\ref{sec:DLmodel}, we train five individual CNN models, and each model is trained with a set of 900 noiseless spectra that we perturb with Gaussian noise (Section~\ref{sec:mock_spectra}). We then use these CNN models to predict the \Lya\ forest parameters of the 15 HIRES quasar spectra from R12 (Section~\ref{sec:observed_data}) and build a catalogue of \Lya\ absorption systems with minimum \Lya ID$>0.3$ for each quasar spectrum. The final determinations of the physical properties (i.e. $\log{N_{\rm{HI}}}$, $z_{\mathrm{HI}}$, and $\log{b_{\mathrm{HI}}}$) are a weighted-average of the predictions for a given absorption line, using \Lya ID as the weights. The mean value of \Lya ID is used as the final probability of a CNN prediction. In order to evaluate the performance of our CNN models, we compare the CNN-classified \Lya\ absorbers with the catalogue built by R12. These authors identified \Lya\ systems and estimated the \HI\ column density, redshift, and Doppler width by Voigt profile fitting. To avoid the proximity effect, \Lya\ systems are excluded if they are within $3000\ \mathrm{km\ s^{-1}}$ of the quasar. Additionally, each \Lya\ system identified by R12 was validated by confirming the existence of at least one other higher order Lyman series transition; when higher order Lyman series lines were available, they were jointly fit. Our network only uses the \Lya\ absorption line.

\subsection{Predicting 15 HIRES observed spectra}
\label{sec:predict_HIRES_spec}
Although our CNN reaches high precision (Equation~\ref{eq:prec_rec}) when predicting fake spectra that only contain \Lya\ absorbers, observed quasar spectra are much more complex and challenging due to the existence of other Lyman series absorption lines and metal lines. Hence, we carry out a post-processing procedure to exclude CNN-classified absorption line systems that: (1) have a redshift greater than the quasar \Lya\ emission redshift; (2) are within a region including higher order Lyman series lines such as \Lyb\ lines; or (3) do not exhibit a \Lyb\ absorption line.

In detail, we first remove the systems with a CNN-estimated redshift larger than or equal to the quasar redshift. To avoid regions including other higher order Lyman series lines, we focus on the region where only \Lya\ absorbers of the Lyman series exist. This is carried out by removing systems located at the wavelengths blueward from the potential highest-redshift \Lyb\ absorber estimated by the quasar emission redshift. Additionally, as in R12, we remove systems that are within $3000\ \mathrm{km\ s^{-1}}$ of the quasar to avoid the proximity effect. Finally, we examine the corresponding \Lyb\ absorption lines for each CNN-classified \Lya\ system using the CNN-predicted redshift, column density, and Doppler width. A CNN-classified \Lya\ system is removed if the following criteria are satisfied: (1) the estimated \Lyb\ flux is much lower than the observed flux, i.e. the difference of fluxes (CNN \Lyb\ flux $-$ observed flux) is negative and its absolute value $>1\sigma$, where $\sigma$ is the median value of the noise spectrum near the centre of the absorption line, defined by the FWHM (i.e. pixels within $\sim\pm12.5\ \mathrm{km\ s^{-1}}$, see Section~\ref{sec:cnn_lya_mockspec}); and (2) the \Lyb\ absorption line is not saturated, i.e. \Lyb\ observed flux $>3\sigma$ (following the definition of saturation in Section~\ref{sec:lya_sys_pixel}).

Additionally, we add two additional flags to our CNN catalouge --- \textit{\Lyb\_inspec\_flag} and \textit{sml\_\Lyb\_flag}. The former decides if the wavelength of a corresponding \Lyb\ absorber of a CNN-classified system is within the observed wavelength range; thus, 1 if yes and 0 if no. The latter flag assesses if the estimated flux of a corresponding \Lyb\ absorber can be hidden within the noise level, i.e. \Lyb\ flux $>(1-\sigma)$. If the \Lyb\ absorption feature can be hidden within $1\sigma$, this flag \textit{sml\_\Lyb\_flag} is set to 1, and the opposite case has \textit{sml\_\Lyb\_flag} $=0$.

\subsection{Comparison with R12 catalogue}
\label{sec:comp_R12}
\begin{figure}{}
\begin{center}
\graphicspath{}
	\includegraphics[width=\columnwidth]{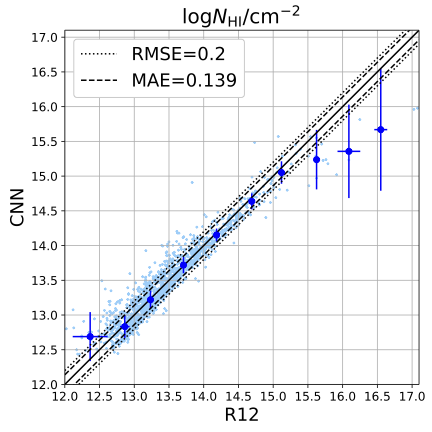}
	\includegraphics[width=\columnwidth]{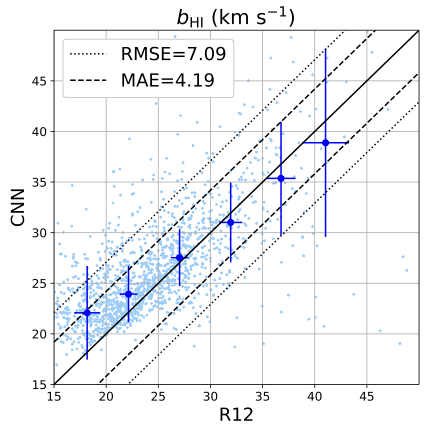}
   	\caption{Comparisons between the CNN and R12 values of $\log{N}_{\mathrm{HI}}/{\rm cm}^{-2}$ (top) and $b_{\mathrm{HI}}$ (bottom). The black solid line shows a one-to-one relation, the dashed lines indicate the scatter defined by the MAE of each plot, and the dotted lines are defined by the RMSE of each plot. Dark blue datapoints show the median values of CNN and R12 within different bins of R12. The $x$-axis error bar is defined by the median value of the estimation errors of the datapoints provided in R12 within different bins of R12, while the $y$-axis error bar presents the MAE of each bin.}
    \label{fig:1to1_compR12_flags}
\end{center}
\end{figure}
\begin{figure*}{}
\begin{center}
\graphicspath{}
	\includegraphics[width=2.1\columnwidth]{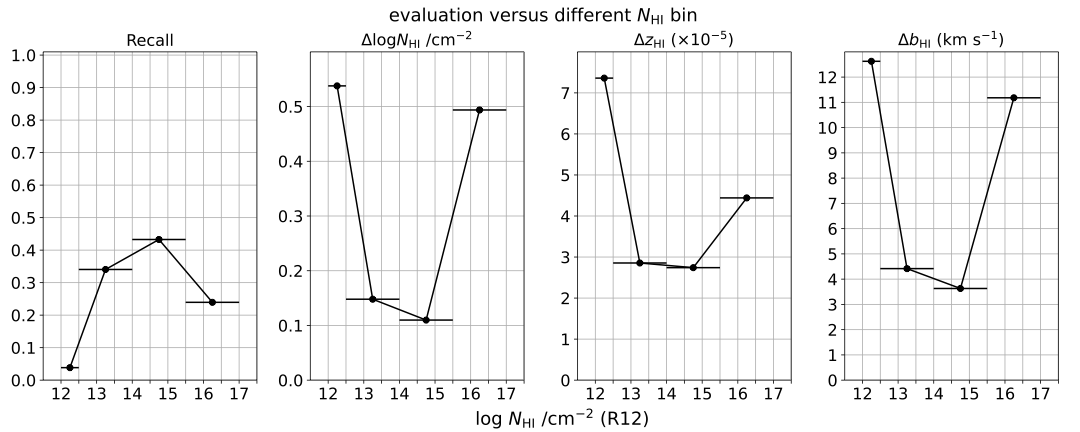}
  	\caption{Evaluation graphs of recall and the MAE of $\log{N_{\mathrm{HI}}}/{\rm cm}^{-2}$, redshift ($z_{\mathrm{HI}}$), and $b_{\mathrm{HI}}$ values, within different $\log{N_{\mathrm{HI}}}/{\rm cm}^{-2}$ bins. The horizontal lines of each datapoint represent the range of each bin; the intervals are 0.5, 1.5, 1.5, and 1.5, from low to high values of $\log{N_{\mathrm{HI}}}/{\rm cm}^{-2}$.}
    \label{fig:obs_eval_diffNHI}
\end{center}
\end{figure*}
To assess the confidence of the CNN results by the R12 catalogue, we focus on the spectral region that only contains \Lya\ absorption lines (Section~\ref{sec:predict_HIRES_spec}). We match the CNN-classified \Lya\ absorbers with the R12 catalogue using the same criteria for simulated spectra described in Section~\ref{sec:cnn_lya_mockspec}, i.e. the velocity difference between the identified systems of our CNN and R12 is smaller than half of the minimum FWHM that can be detected by our machine: $\sim12.5\ \mathrm{km\ s^{-1}}$.

Since the R12 catalogue is based on a consistent fit of all available Lyman series lines (i.e. \Lya\ and at least one other Lyman series transition), a mismatch between R12 and our CNN could happen if: (1) the corresponding \Lyb\ absorber of a CNN-classified system is out of the wavelength range of the spectrum, or (2) weak \Lyb\ absorption lines that are buried within the noise level of a broad absorption feature. By applying two additional flags\footnote{Without applying the two flags, one can just compare the systems that are within the same redshift ranges as the ones that were fit in R12. 
The precision, recall, and MAE of $\Delta\log{N_{\mathrm{HI}}}/{\rm cm}^{-2}$, $\Delta{z_{\mathrm{HI}}}$, and $\Delta{b_{\mathrm{HI}}}$ are 0.778, 0.284, 0.14 dex, 2.7$\times10^{-5}$, and 4.3~${\rm km~s^{-1}}$, respectively.}: (1) \textit{\Lyb\_inspec\_flag} $=1$ and (2) \textit{sml\_\Lyb\_flag} $=0$, $\sim78$ per cent of the ML-classified \Lya\ systems are matched with R12, i.e. precision $=0.78$.\footnote{It may be possible to include \Lyb\ lines in the training process. This may improve the precision of the predictions of the \Lya\ lines. We leave this as an exercise for future work.} The comparison of column density ($\log{N_{\mathrm{HI}}}/{\rm cm}^{-2}$) and Doppler width ($b_{\mathrm{HI}}$) between our CNN and R12 are shown in Fig.~\ref{fig:1to1_compR12_flags} (also check Table~\ref{tab:evaluation_diffsets}). Dark blue datapoints are the median values of each bin of R12. The bin interval is 0.5 for $\log{N_{\mathrm{HI}}}/{\rm cm}^{-2}$ and 5 $\mathrm{km\ s^{-1}}$ for $b_{\mathrm{HI}}$. There is a statistical uncertainty associated with each quantity in the R12 catalogue based on Voigt profile fitting; the $x$-axis error bar uses the median value of the deviations to represent the typical error of each quantity in R12 within different bins. On the other hand, the $y$-axis error bar presents the MAE of the datapoints in each bin.

Compared to the simulated spectra results in Fig.~\ref{fig:mock_N_b_1to1}, the CNN performance decreases when predicting real spectra. This is due to the more complex blending of features that are seen in observational data. As discussed in Section~\ref{sec:mock_FN}, our CNN tends to use only one line to recover a broad absorption feature while it is generally composed of multiple neighbouring lines. In this case, even though there is a matched \Lya\ system between the two catalogues, the CNN predictions of $\log{N_{\mathrm{HI}}}$ and $b_{\mathrm{HI}}$ will not be consistent with the values listed in the R12 catalogue.

Nevertheless, our CNN models do a good job in predicting \HI\ column density $\log{N_{\mathrm{HI}}}/{\rm cm}^{-2}$ with MAE$=0.139$. In particular, the range between $12.5\leq\log{N_{\mathrm{HI}}}/\mathrm{cm^{-2}}<15.5$ shows a tight one-to-one relation with MAE$=0.135$ (also see Table~\ref{tab:evaluation_diffsets}). Outside this column density range, the number of CNN-classified \Lya\ systems are much fewer (42 out of 1\,930 CNN-classified \Lya\ systems) which results in a larger scatter within this range. This indicates that our CNN has difficulty in correctly classifying these absorption lines and leads to a relatively poor estimate of the \HI\ column density for \Lya\ systems with $\log{N_{\mathrm{HI}}}/{\rm cm}^{-2}<12.5$ or $\log{N_{\mathrm{HI}}}/{\rm cm}^{-2}>15.5$. Note that the R12 data are of considerably higher S/N compared to the simulated data that were used to train our CNN model. As described in Section~\ref{sec:mock_FN}, weak low \HI\ column density systems are often buried in noise near the continuum level for data of S/N $\simeq10$. To improve the prediction of lower \HI\ column density systems, one may train a model with higher S/N input spectra, and apply this model to observational data of comparably high S/N (see Appendix~\ref{app:diff_snr_R12}).

The catalogue comparison of $b_{\mathrm{HI}}$ (bottom panel of Fig.~\ref{fig:1to1_compR12_flags}) shows a similar trend to the results of simulated spectra in Fig.~\ref{fig:mock_N_b_1to1} with a larger scatter. Within the range of $b_{\mathrm{HI}}<20~{\rm km~s}^{-1}$, our CNN tends to overestimate the $b_{\mathrm{HI}}$ value, because the CNN models use a single \Lya\ line to recover a feature that is composed of multiple lines. On the other hand, for larger $b_{\mathrm{HI}}$ values, our CNN has difficulty to predict systems with broader absorption features due to the restriction of the scanning window size (see point (iii) of the false negative summary in Section~\ref{sec:mock_FN}). The window size is a hyperparameter tuned to optimise the predictions for the majority of \Lya\ absorbers (Section~\ref{sec:cnn_design}). There are $\sim93\%$ of \Lya\ absorbers with $b_{\mathrm{HI}}<35~{\rm km~s}^{-1}$ from R12 ($\sim97\%$ of them with $b_{\mathrm{HI}}<40~{\rm km~s}^{-1}$), and the predictions for the systems with a larger $b_{\mathrm{HI}}$ value are worse (this also occurred for the simulated spectra).

Finally, for different column density bins (defined using the R12 catalogue), we present the recall and the MAE of $\log{N_{\mathrm{HI}}}/{\rm cm}^{-2}$, redshift ($z_{\mathrm{HI}}$), and $b_{\mathrm{HI}}$ in Fig.~\ref{fig:obs_eval_diffNHI}. Based on the column density comparison shown in Fig.~\ref{fig:1to1_compR12_flags}, we separate matched samples into four bins: $\log{N_{\mathrm{HI}}}/{\rm cm}^{-2}=12.0-12.5$, $12.5-14.0$, $14.0-15.5$, and $15.5-17.0$. Fig.~\ref{fig:obs_eval_diffNHI} demonstrates that the CNN predictions of different physical properties are most consistent with R12 within the \HI\ column density range $12.5\leq\log{N_{\mathrm{HI}}}/{\rm cm}^{-2}<15.5\ \mathrm{cm^{-2}}$ (Table~\ref{tab:evaluation_diffsets}).

Note that the CNN-classified \Lya\ absorbers for the above results are identified by at least one CNN model out of five models. To further improve the CNN predictions, one can impose a selection criterion to the number of the CNN models that identify a \Lya\ system. For example, by requiring that a \Lya\ absorber must be identified by at least two CNN models, the precision increases from 0.78 to 0.85, and the overall MAE for $\log{N_{\mathrm{HI}}}/{\rm cm}^{-2}$ improves slightly, and the MAE for $b_{\mathrm{HI}}$ drops to $3.8\ \mathrm{km\ s^{-1}}$, respectively (see the results of other test datasets in Table~\ref{tab:evaluation_diffsets}).

\subsubsection{False positive and false negative}
\label{sec:obsFPs_FNs}
\begin{figure*}{}
\begin{center}
\graphicspath{}
	\includegraphics[width=2.1\columnwidth]{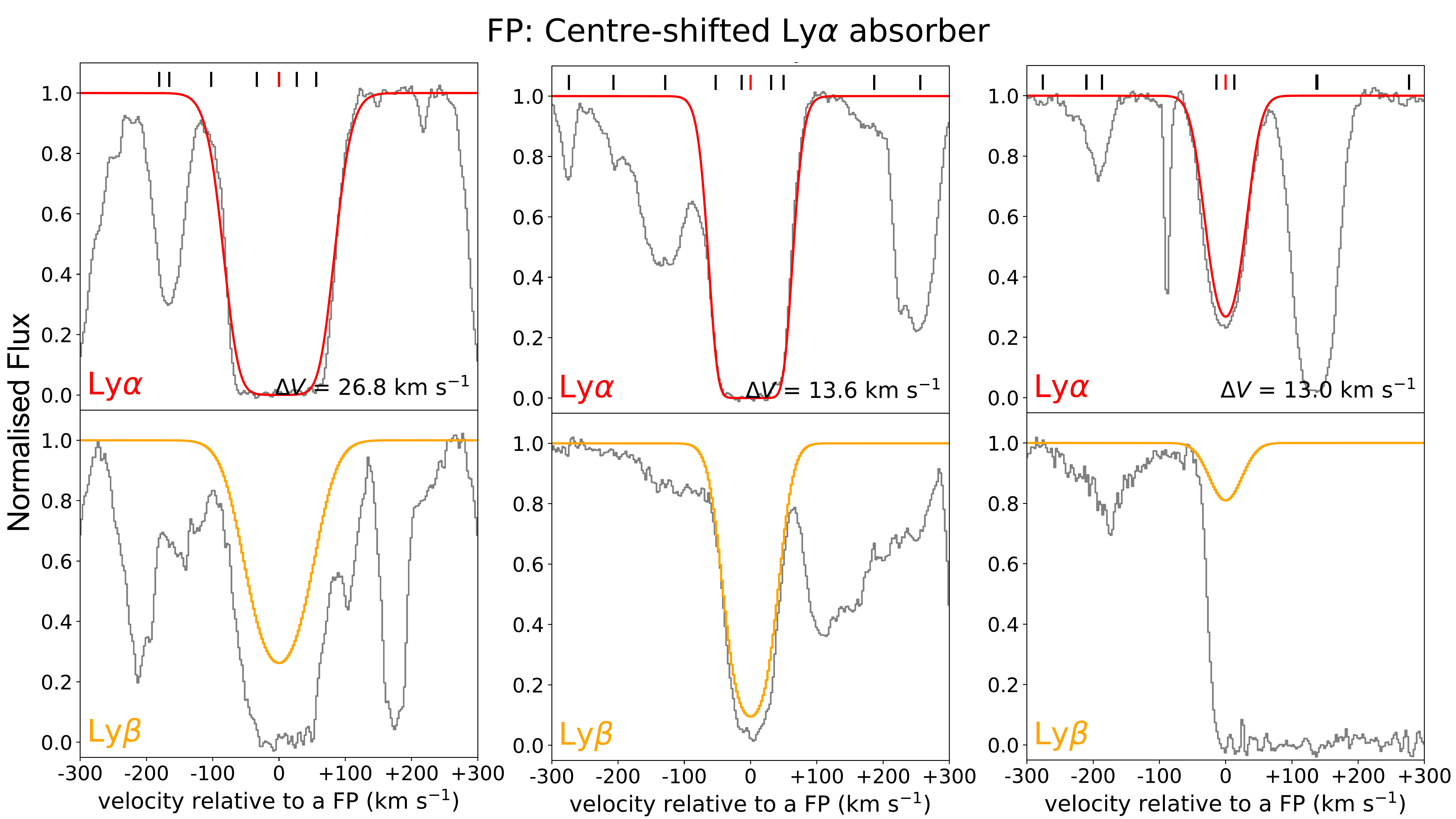}
  	\caption{Examples of potential \Lya\ absorbers (red curves; top panels) that are not matched with an absorber in R12 due to multiple components in R12 being fit with a single absorber by the CNN. Due to velocity differences between the two catalogs, they are flagged as false positives although the absorber is identified in both catalogs. The centre of the FP is labelled by a red short tick mark. These examples of FPs are due to the matching criterion (see Section~\ref{sec:mock_FP}). The orange curves in the bottom panels show the corresponding \Lyb\ absorber. The gray histograms show the observational data. Black short tick marks above the spectra indicate the centres of the \Lya\ systems from R12. The $\Delta{V}$ values shown in the top panels provide the velocity difference between the prediction and the nearest absorption line from the catalogue of R12.}
    \label{fig:obs_FP_example_mismatch}
\end{center}
\end{figure*}
\begin{figure}{}
\begin{center}
\graphicspath{}
	\includegraphics[width=0.8\columnwidth]{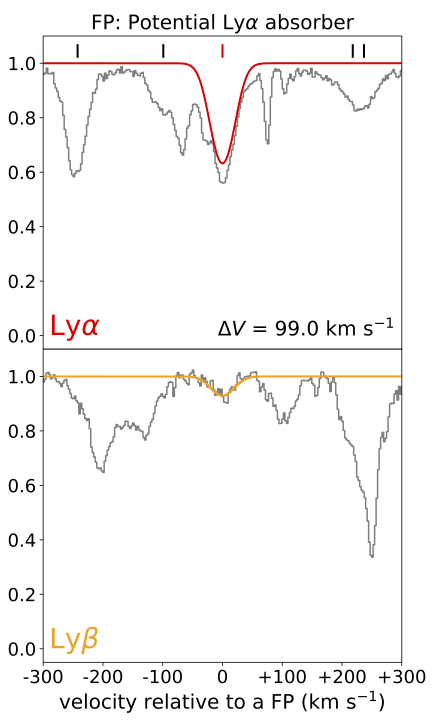}
  	\caption{Same as Fig.~\ref{fig:obs_FP_example_mismatch}, but shows an example of a potential \Lya\ absorber that is not listed in the R12 catalogue.}
    \label{fig:obs_FP_example_potentialLya}
\end{center}
\end{figure}
Except for misclassification, which is dominated by contaminating metal lines, one of the primary causes of false positives in the observational data is due to the velocity threshold ($\sim12.5\ \mathrm{km\ s^{-1}}$) used to match systems between the CNN and R12 catalogues (as discussed in Section~\ref{sec:mock_FP}). Fig.~\ref{fig:obs_FP_example_mismatch} shows three examples of this FP case. The CNN tends to fit a broad absorption feature with one line, while it is composed of multiple neighbouring lines in the R12 catalogue. We notice that the broader an absorption feature is, the worse CNN predictions are obtained, e.g., the leftmost panel in Fig.~\ref{fig:obs_FP_example_mismatch}. Additionally, since the classifications in R12 might have missed some \Lya\ systems from manual Voigt profile fitting, some FPs by our CNN could be a potential \Lya\ absorber. Examples are shown in Fig.~\ref{fig:obs_FP_example_potentialLya}. We compared the CNN identified systems with the robust \Lya\ absorbers from R12 which were validated with higher order lines, e.g., \Lyb, Ly$\gamma$, etc. Hence, there may be mismatch because the corresponding \Lyb\ or Ly$\gamma$ lines of a potential \Lya\ absorber is difficult to detect. For example, in Fig.~\ref{fig:obs_FP_example_potentialLya}, we show an example with possible \Lya\ and \Lyb\ absorption consistent with a true absorption system. However, this example may instead be due to a metal line absorption line, given that there are several neighbouring metal line absorbers nearby.

\begin{figure}{}
\begin{center}
\graphicspath{}
	\includegraphics[width=\columnwidth]{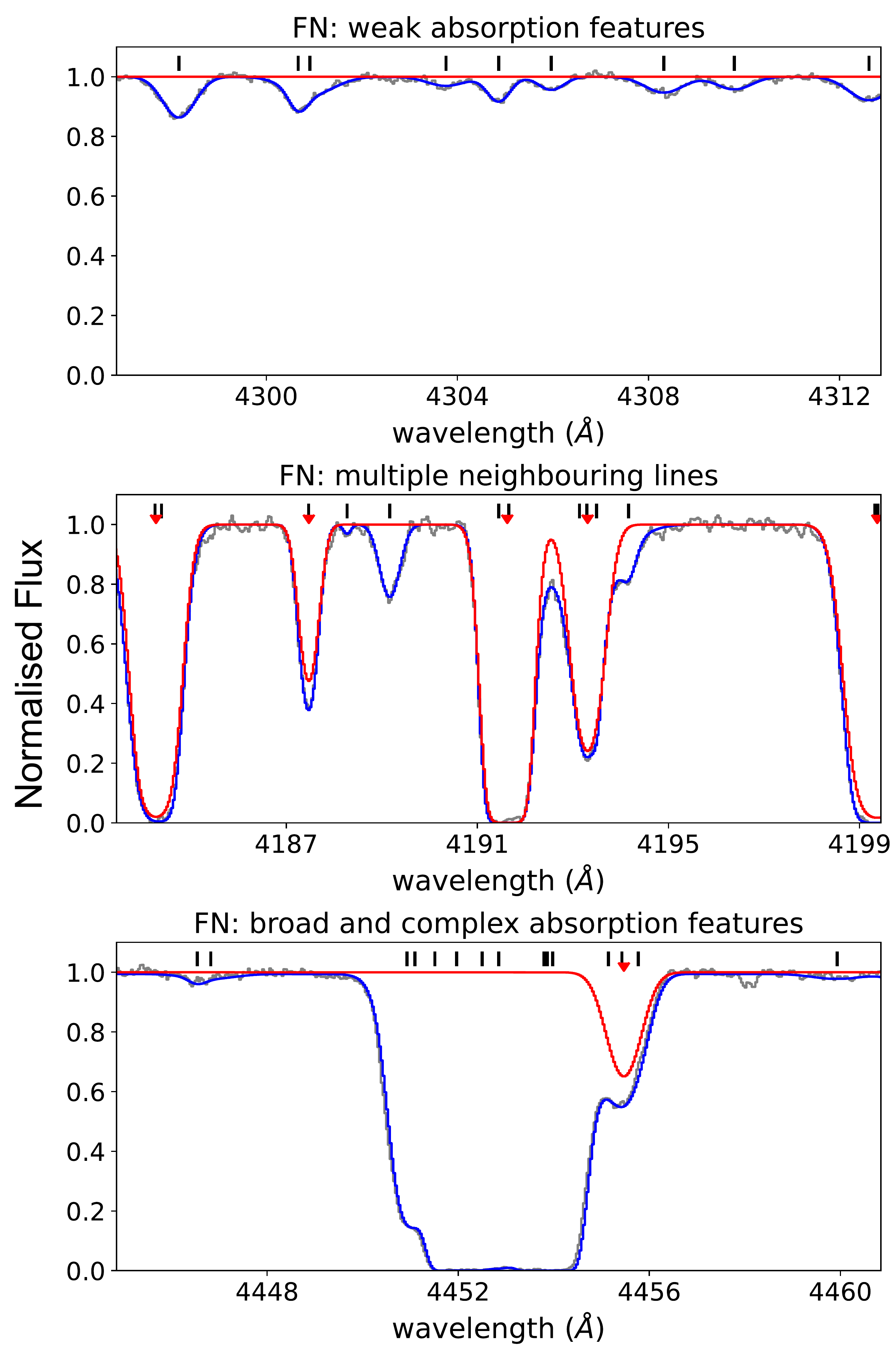}
  	\caption{Several examples of false negatives. The centres of each \Lya\ system from R12 are labelled by black short tick marks above the spectra, while the red arrow indicates a match between the CNN and R12 results. From the top to bottom panel, we showcase different reasons responsible for false negatives. The gray curve represents the data, the blue curves represent a reconstruction of the data based on the \Lya\ absorbers listed in the R12 catalogue, and the red curve is a reconstruction based on all of the predicted lines by our CNN.}
    \label{fig:obs_FN}
\end{center}
\end{figure}
The false negatives identified with the observational data are contributed by the same sources as the ones discussed in Section~\ref{sec:mock_FN} (i.e. low column density absorption features that are buried in the noise). As mentioned in Section~\ref{sec:mock_FN}, our choice to train a model on low S/N data is a compromise to allow a CNN technique to be applied to spectra with both low and high S/N. For completeness, we have also trained a model with S/N closer to the quasar spectra from R12 and we test this model using observed spectra. This comparison is discussed in Appendix~\ref{app:diff_snr_R12}.

In Fig.~\ref{fig:obs_FN} we showcase examples of the different cases of FNs. The top panel presents the case of FNs having weak absorption features that are missed by our CNN, which is trained with noisy spectra. In the middle panel, our CNN uses one \Lya\ line to describe the absorption feature, while there are multiple nearby lines listed in R12 responsible for this feature. This specific case also contributes a false positive depending on the distance between a CNN-classified system and its closest system from R12. Finally, we showcase a broad and complex absorption feature containing many \Lya\ absorbers in the bottom panel. Our CNN has difficulty analysing a broader feature such as the one showcased here, since our CNN is trained with only primary lines. When an absorption feature is broader than the structure that our CNN can reconstruct with one \Lya\ line, the CNN fails to classify.

Since the behaviour of false negatives using our CNN can be determined empirically, a correction factor can be calculated to convert the predicted distribution of \Lya\ forest absorbers to the intrinsic (i.e. input) distribution of \Lya\ absorbers. We will consider this approach in a future paper.

\subsubsection{Predicting HIRES spectra with different S/N}
\label{sec:HIRES_diffSNR}
\begin{figure*}{}
\begin{center}
\graphicspath{}
	\includegraphics[width=2.1\columnwidth]{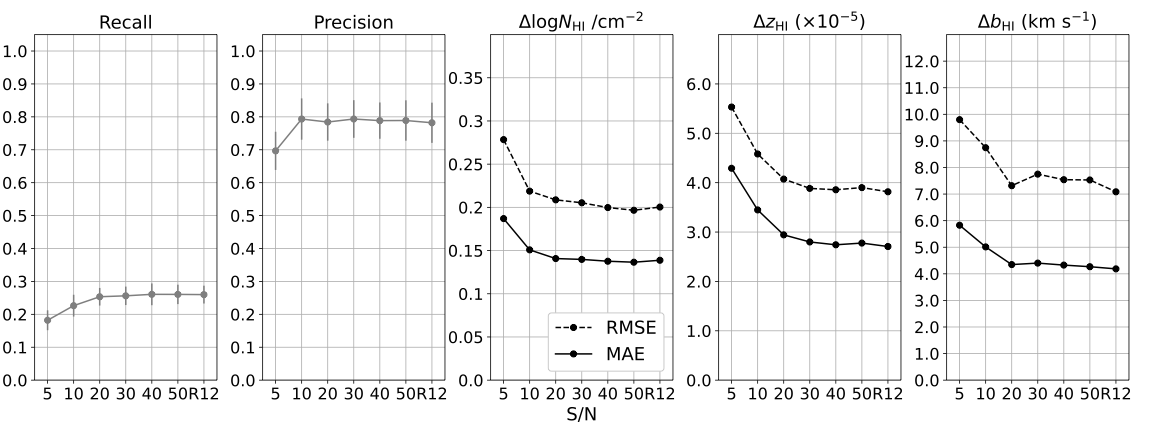}
   	\caption{Our CNN predictions of the R12 HIRES spectra are stable when we artificially degrade the R12 data. Note that the `R12' data points represent the predictions to the high quality HIRES spectra from R12 (S/N~$\gtrsim50$). From left to right, we show the recall, precision, the RMSE (black dashed line) and MAE (black solid line) of \HI\ column density ($\Delta \log{N_{\mathrm{HI}}}/{\rm cm}^{-2}$), redshift ($\Delta z_{\mathrm{HI}}$), and the Doppler width ($\Delta b_{\mathrm{HI}}$).}
    \label{fig:HIRES_diffSNR}
\end{center}
\end{figure*}
In this section, we test if our CNN is capable of predicting observed Keck/HIRES spectra of different S/N. Additional noise is added to the high quality HIRES spectra from R12 to degrade the S/N. We test different cases from S/N $=5$ to S/N $=50$ (i.e. the latter case represents the lower S/N end of the R12 spectra). This test is to ensure that in future works we can further apply our trained CNN models to predict spectra in the HIRES archives such as the Keck Observatory Database of Ionized Absorption toward Quasars (KODIAQ) survey \citep{OMeara2017, OMeara2021}. Fig.~\ref{fig:HIRES_diffSNR} demonstrates that the performance of our CNN is consistent with the predictions of simulated spectra (see Appendix~\ref{app:diff_snr}). This again confirms that training the CNN with noisy spectra is of great importance to stabilise the predictions of spectra with different noise levels. Although there is a drop in the CNN performance at S/N $<20$, the changes are still within an acceptable range for further scientific analyses. By training and testing a CNN applied to high redshift quasar spectra, we have opened up the possibility to efficiently and effectively harvest the information buried in the \Lya\ forest. This is an important step towards understanding and analysing the significant amount of data that will be acquired with future facilities.

\section{Summary}
\label{sec:summary}
We have developed a machine learning based detection algorithm using convolutional neural networks (CNN) to derive the physical parameters of \Lya\ absorbers within the forest of high-resolution QSO absorption line spectra. In particular, we focus on the low \HI\ column density systems ($N_{\mathrm{HI}}$ $<10^{17}\ \mathrm{cm^{-2}}$) and predict their physical properties such as \HI\ column density ($\log{N}_{\mathrm{HI}}/{\rm cm}^{-2}$), redshift ($z_{\mathrm{HI}}$), and Doppler width ($b_{\mathrm{HI}}$). The low column density \Lya\ absorbers serve as a great tracer to the thermal history of the low-density IGM and can be used to probe the baryonic matter distribution. However, since they can be easily contaminated by other Lyman series and metal lines, previous applications of machine learning to the \Lya\ forest have focused on identifying DLAs ($N_{\mathrm{HI}}$ $\ge10^{20.3}\ \mathrm{cm^{-2}}$) which show strong, damped absorption features.

Our CNN model is trained with 900 noisy simulated spectra with a S/N drawn from a Gaussian distribution of mean $=10$ and standard deviation $=2$. This training strategy stabilises the CNN performance when predicting spectra of different S/N (Appendix~\ref{app:diff_snr} and \ref{app:diff_snr_R12}) and allows us to apply our CNN models to the current archives of spectroscopic data, as well as future surveys. The simulated spectra that we use for training our model represent quasars at redshift $z=3$ and are convolved with an instrumental resolution of $v_{\rm FWHM}=7\ \mathrm{km\ s^{-1}}$. These values are typical of the data in current observatory archives. Different FWHM values have no impact on the performance of the CNN model (i.e. the \Lya\ forest absorption lines are fully resolved), while at higher redshifts there is increased blending due to neighbouring absorption features, which negatively impacts the accuracy of the CNN predictions (see Appendix~\ref{app:diff_spec}). The pixel size of the simulated spectra is set to $2.5\ \mathrm{km\ s^{-1}}$.

We first examine the CNN performance with simulated spectra, and match the CNN prediction and true systems using a velocity threshold defined by half of the minimum FWHM $\sim12.5\ \mathrm{km\ s^{-1}}$ (estimated by $b_{\mathrm{HI}}=15\ \mathrm{km\ s^{-1}}$). By matching the predicted systems with the systems listed in the true catalogue, over 99\% of the CNN-classified \Lya\ systems are true. However, the completeness is low ($\sim 32\%$), i.e. only a small fraction of the \Lya\ systems are identified by our CNN. We summarise three types of false negative: (1) weak absorption features that might be neglected by our CNN due to the limitation of the noisy training spectra; (2) a strong absorption feature composed of multiple neighbouring lines, contributing one false positive and many false negatives; (3) broad and complex absorption features that cannot be represented by one \Lya\ absorber. Case (1) dominates the FN; the completeness increases to 77\% when excluding this case of FN.

We then train five individual CNN models to predict 15 HIRES spectra and compare the CNN predictions with the results of manual Voigt profile fitting by R12. While the manual method costs 1-2 years for the 15 spectra in R12, the prediction process by our CNN costs less than three minutes per quasar spectrum with a size of $\sim120\,000$ pixels using a MacBook Pro with a 2.3 GHz Intel Core i7 processor and Intel Iris Plus Graphics 1536 MB.

Since an observed spectrum contains complex structures and contamination such as metal lines, a post-processing procedure is carried out to exclude unreliable ML-classified \Lya\ systems. Around 78 percent of ML-classified \Lya\ systems are matched with R12. There are three sources responsible for false positives: (1) a simple mismatch due to the chosen velocity threshold ($\sim12.5\ \mathrm{km\ s^{-1}}$); (2) a potential \Lya\ absorber that is not listed in R12 due to weak or hidden \Lyb\ absorption; and (3) misclassification due to broad absorption features formed by multiple blended metal lines. We further conclude that the CNN models provide the most reliable predictions within the range of $12.5\leq\log{N_{\mathrm{HI}}}/\mathrm{cm^{-2}}<15.5$. Within this range, the MAE of $\log{N}_{\mathrm{HI}}/\mathrm{cm^{-2}}$, $z_{\mathrm{HI}}$, and $b_{\mathrm{HI}}$ are $0.13$~dex, $2.7\times{10}^{-5}$, and $4.1\ \mathrm{km\ s^{-1}}$, respectively, demonstrating the accuracy of our CNN predictions. We conclude that a general-purpose CNN applied to the \Lya\ forest may not be as effective as one that is trained for a specific science goal, and it is important to better understand the parameter space where a model succeeds or under performs. We found that the false negatives occur under the same conditions for both simulated and observed spectra.

Although we train the CNN models with noisy (S/N $\simeq10$) simulated spectra, they provide consistent performance when predicting much higher quality (S/N $\gtrsim70$) observational spectra. This gives us confidence that our model can be applied to both cosmological simulations and observations of the \Lya\ forest, and help to provide an insight into some of the missing ingredients in simulations.

Finally, we examine the CNN performance when predicting observed Keck/HIRES spectra of different S/N, and draw the same conclusions as the analysis of the simulated spectra. An investigation can be further carried out to quantify the impact of different S/N on the `accuracy' of the conventional analyses to observed spectra. More importantly, this result validates the possibility to apply a CNN model with our approach to analyse the enormous quantity of data that will be obtained with future facilities.



\section*{Acknowledgements}
We thank an anonymous referee for a timely and thorough report that helped to clarify various aspects of the paper. T.-Y. Cheng acknowledges the support of STFC grant ST/T000244/1 and Royal Society grant RF/ERE/210326, hosted at Durham University, and the support by Towards Turing 2.0 under the EPSRC Grant EP/W037211/1 \& The Alan Turing Institute.
During this work, RJC was supported by a Royal Society University Research Fellowship. RJC acknowledges support from STFC (ST/T000244/1).



\section*{Data Availability}
The observed Keck/HIRES spectra are publicly available on the Keck Observatory Archive. The machine learning code is not published, but may be shared upon request.

\bibliographystyle{mnras}
\bibliography{ms}

\appendix

\section{The impact on predicting spectra with different initial setups}
\label{app:diff_spec}
\begin{figure*}{}
\begin{center}
\graphicspath{}
	\includegraphics[width=2.1\columnwidth]{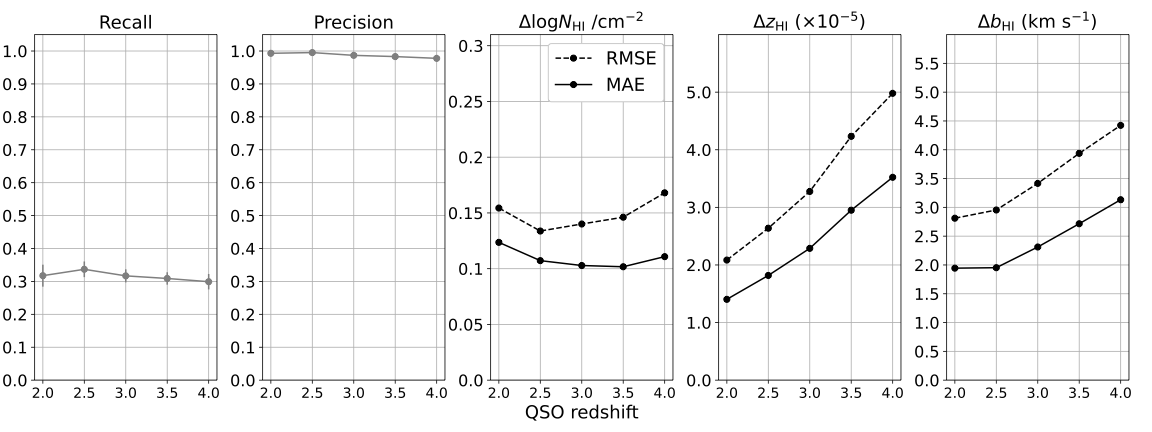}
	\includegraphics[width=2.1\columnwidth]{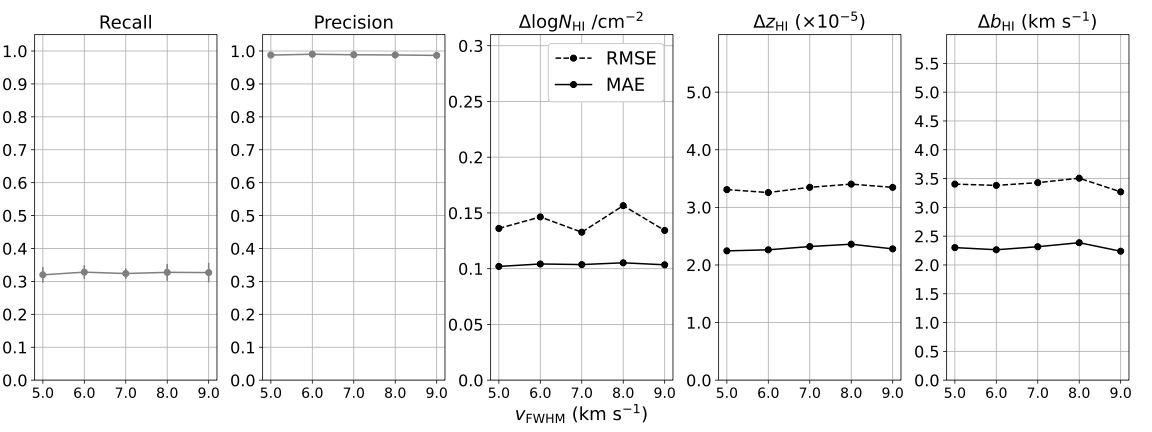}
	\includegraphics[width=2.1\columnwidth]{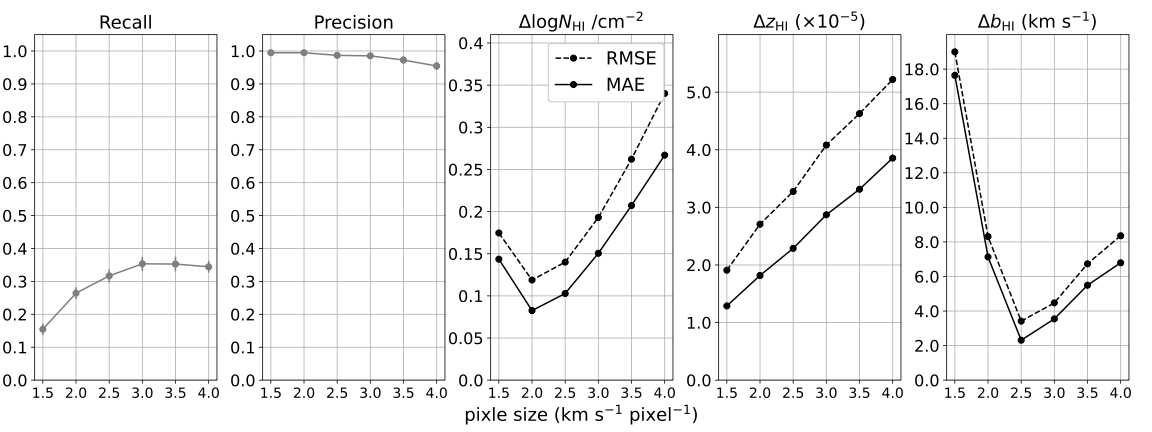}
   	\caption{The overall performance of our CNN on noiseless spectra with different observing setups. We illustrate the sensitivity to quasar redshift (top row), instrumental FWHM resolution (middle row), and pixel size (bottom row). Note that the CNN model is trained using simulated noisy quasar spectra at redshift $z=3$ and are observed using an instrumental FWHM resolution of $v_{\rm FWHM}=7\ \mathrm{km\ s^{-1}}$ and sampled with vpix$=2.5\ \mathrm{km\ s^{-1}\ pixel^{-1}}$ (Section~\ref{sec:mock_spectra}). From left to right, the panels show the recall, precision, RMSE (black dashed line) and MAE (black solid line) of \HI\ column density ($\Delta \log{N_{\mathrm{HI}}}/{\rm cm}^{-2}$), redshift ($\Delta z_{\mathrm{HI}}$), and the Doppler width ($\Delta b_{\mathrm{HI}}$).}
    \label{figapp:diffspec}
\end{center}
\end{figure*}
We test our pre-trained CNN model on 25 newly generated spectra with: (1) quasars at different redshifts; (2) different instrument resolution (FWHM); and (3) data that are sampled with different pixel size (vpix). This test is to validate the feasibility of our CNN model to predict observational spectra that were acquired with different setups.

Firstly, for different quasar redshifts (top row in Fig.~\ref{figapp:diffspec}), the recall and precision remain consistent and the $\log{N_{\mathrm{HI}}}/{\rm cm}^{-2}$ prediction does not show significant deviation. However, the RMSE and MAE of $z_{\mathrm{HI}}$ and $b_{\mathrm{HI}}$ estimations increase as a quasar emission redshift increases. This means that the prediction accuracy decreases as a quasar redshift increases. This result is caused by the increased blending due to \Lya\ forest absorption lines at higher redshift. During the early Universe, \HI\ gas clouds are more abundant and their absorption features overlap in velocity space. This overlap introduces additional uncertainty in the predicted physical properties; this is true for both a machine-learning based algorithm or conventional Voigt profile fitting. Even though the RMSE and MAE of $z_{\mathrm{HI}}$ and $b_{\mathrm{HI}}$ are within a factor of $\sim1.5$ of the RMSE and MAE at $z\simeq3$, we conclude that a CNN tailored to a specific redshift may further improve the results, depending on the science application.

In the middle row of Fig.~\ref{figapp:diffspec}, we generated spectra with different instrument FWHM resolution over a narrow range 5 < FWHM/(km/s) < 9, which samples the relevant resolutions of current high dispersion spectrographs like Keck/HIRES and ESO/UVES \citep[European Southern Observatory Ultraviolet and Visual Echelle Spectrograph;][]{Dekker2000}. Overall, different FWHMs in this range show no impact to both the detection and physical property estimates. This is because the widths of the \Lya\ forest absorption lines (FWHM$\gtrsim15~{\rm km~s}^{-1}$) are usually fully resolved at the instrument resolution of typical spectrographs, such as Keck/HIRES and VLT/UVES (FWHM$\simeq7~{\rm km~s}^{-1}$). Finally, in the bottom row of Fig.~\ref{figapp:diffspec}, we show that the choice of pixel size introduces a serious impact to the results, in particular the Doppler width, $b_{\mathrm{HI}}$. However, this issue can be circumvented by resampling. If an input spectrum is not sampled with vpix$=2.5~{\rm km~s}^{-1}$, we resample the input data to ensure that our CNN model produces reliable results. This resampling process does not impact the trained network, nor the results, since the \Lya\ forest absorption lines are fully resolved.

\section{Bayesian optimisation}
\label{app:bayesian_opt}
The predictions by a network strongly depend on its hyperparameters such as the number of neurons, dropout rate, the kernel sizes, etc. A failed prediction of a network might be simply due to an unoptimised architecture used for training. Hence, it is of great importance to select a set of hyperparameters that provide the most optimal combination for a specific goal. This selection is often done by a brute-force method -- grid searches -- such that all possible combinations of each hyperparameter are evaluated. This method is therefore extremely time-consuming, and the tested sets of hyperparameters are limited due to computational allowance.

Unlike grid searches, Bayesian optimisation \citep{Snoek2012} provides a `smart guess' to approach an optimal combination of hyperparameters: $x_1, x_2, ..., x_n$, where $n$ represents the number of hyperparameters. This process is much faster than the grid searches to find a set of hyperparameters that performs well. The concept is to model the network's function $f\left( x_1, x_2, ..., x_n \right)$ to a surrogate analytical function. In this work, we use a Gaussian process \citep{Rasmussen2006} which forms the prior distribution as multivariate normal distributions. As providing data, the posterior probability distribution $f$ given $f\left( x_1, x_2, ..., x_n \right)$ is computed, and approaches to the prior using a chosen acquisition function which we use the default function $-$ Expected Improvement \citep{Jones1998}. A detailed tutorial is described in \citet{Frazier2018}.

Our optimisation process uses \textsc{GPyOpt} \citep{gpyopt2016}\footnote{\url{http://sheffieldml.github.io/GPyOpt/}} by running 60 iterations to search for the optimal set. A final set of hyperparameters for the model trained with noisy quasar spectra (where the S/N is drawn from a Gaussian distribution of a mean $=10$ and a standard deviation $=2$) is listed in Table~\ref{tab:hyper-parameters}.

\section{The impact of different S/N on mock spectra}
\label{app:diff_snr}
\begin{figure*}{}
\begin{center}
\graphicspath{}
	\includegraphics[width=2.1\columnwidth]{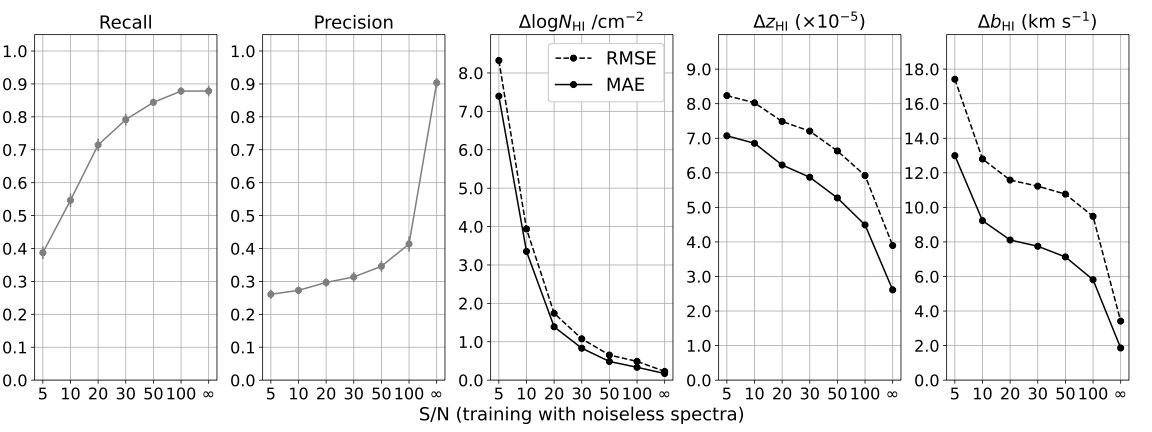}
	\includegraphics[width=2.1\columnwidth]{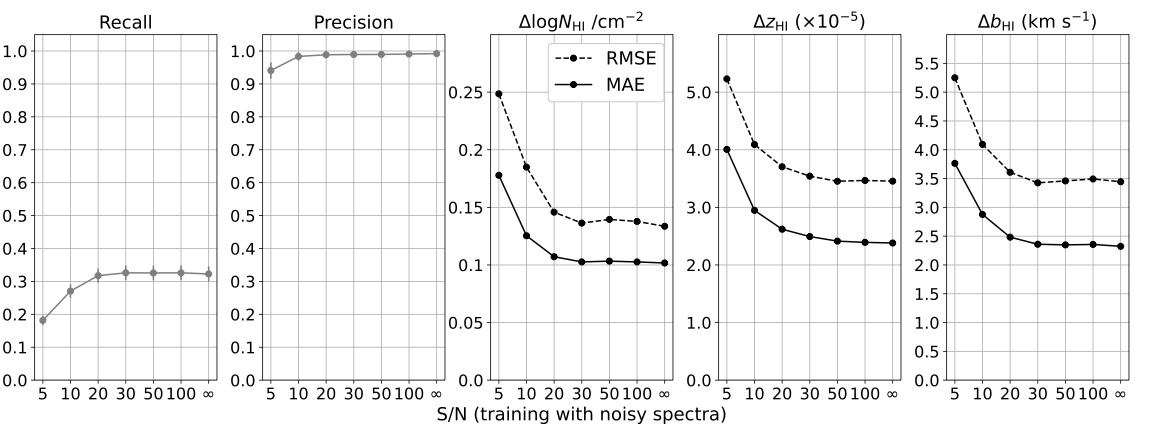}
   	\caption{CNN analysis of spectra with different S/N using two CNN models trained with noiseless spectra (top row) and a Gaussian distribution of S/N with a mean $=10$ and a standard deviation $=2$ (bottom row), respectively. From left to right, we present the recall, precision, the RMSE (black dashed line) and MAE (black solid line) of \HI\ column density ($ \Delta \log{N_{\mathrm{HI}}}/{\rm cm}^{-2}$), redshift ($\Delta z_{\mathrm{HI}}$), and the Doppler width ($\Delta b_{\mathrm{HI}}$). Note that all metrics in the bottom row show a high level of stability for S/N $>20$, demonstrating the more general success of training a CNN model with somewhat low S/N data.}
    \label{figapp:diffSNR}
\end{center}
\end{figure*}
\begin{table}
	\centering
	\begin{tabular}{llc}
		\hline
		\multicolumn{1}{c}{} & {Hyperparameters} & {Optimised value}\\
		\hline\hline
		\multicolumn{1}{l}{Data Input} & {window size ($ws$)} & {259} \\
		\multicolumn{1}{l}{} & {$cnpix$} & {5} \\
		\hline
		\multicolumn{1}{l}{CNN} & {L2} & {0.0} \\
		\multicolumn{1}{l}{Architecture} & {dropout} & {0.0} \\
		\multicolumn{1}{l}{} & {conv\_filter\_1} & {128} \\
		\multicolumn{1}{l}{} & {conv\_filter\_2} & {128} \\
		\multicolumn{1}{l}{} & {conv\_filter\_3} & {128} \\
		\multicolumn{1}{l}{} & {conv\_kernel\_1} & {8} \\
		\multicolumn{1}{l}{} & {conv\_kernel\_2} & {7} \\
		\multicolumn{1}{l}{} & {conv\_kernel\_3} & {8} \\
		\multicolumn{1}{l}{} & {dense\_1} & {128} \\
		\multicolumn{1}{l}{} & {dense\_2\_ID} & {256} \\
		\multicolumn{1}{l}{} & {dense\_2\_N} & {512} \\
		\multicolumn{1}{l}{} & {dense\_2\_z} & {256} \\
		\multicolumn{1}{l}{} & {dense\_2\_b} & {128} \\
		\hline
	\end{tabular}
	\caption{Hyperparameters used in a CNN architecture trained with noiseless spectra. These values are selected using a Bayesian optimisation algorithm.}
	\label{apptab:hyper-parameters}
\end{table}
We tested the impact of training a CNN model using spectra of different noise levels. In Fig.~\ref{figapp:diffSNR}, we show the results of training a CNN model with noiseless spectra (top; hyperparameters used in the CNN architecture are shown in Table~\ref{apptab:hyper-parameters}) and noisy spectra (bottom; Table~\ref{tab:hyper-parameters}) with S/N drawn from a Gaussian distribution with a mean of $10$ and a standard deviation of $2$ (Section~\ref{sec:data_input}). This figure clearly shows that a model trained with noiseless spectra cannot be used to predict new spectra with even a modest amount of noise. Although this model can reach a higher overall recall for noiseless data, the precision and the parameter determinations of $\log{N_{\mathrm{HI}}}/{\rm cm}^{-2}$, $z_{\mathrm{HI}}$, and $b_{\mathrm{HI}}$ are poorly known when noise is added to the testing spectra.

Compared to this, a CNN model trained with spectra involving a distribution of S/N shows stable performance when predicting spectra with different noise levels. A drop in performance occurs to spectra with S/N $<20$. Testing on spectra with S/N $=10$, there is a drop in recall which does not decrease the precision. This indicates that many true \Lya\ absorbers might be hidden in the noise, and our CNN has difficulty to identify them. However, $\sim98\%$ of the CNN-classified \Lya\ systems are classified correctly compared with the list of true systems.

The precision then drops to $\sim0.94$ when analysing spectra of S/N $=5$, and there are significant changes to the RMSE and MAE for the estimates of the physical properties. We did not expect our CNN model to perform well when analysing spectra with S/N $=5$, since this noise level is beyond the range included in our training spectra. However, the changes to the predictions are minor compared to the top panels of Fig.~\ref{figapp:diffSNR} using the model trained with noiseless spectra. Additionally, they are still within an acceptable range for scientific analyses. Hence, with caution, this CNN model can be used to analyse spectra with S/N $>5$.

\section{The impact of different S/N on R12 spectra}
\label{app:diff_snr_R12}
\begin{figure*}{}
\begin{center}
\graphicspath{}
	\includegraphics[width=2.1\columnwidth]{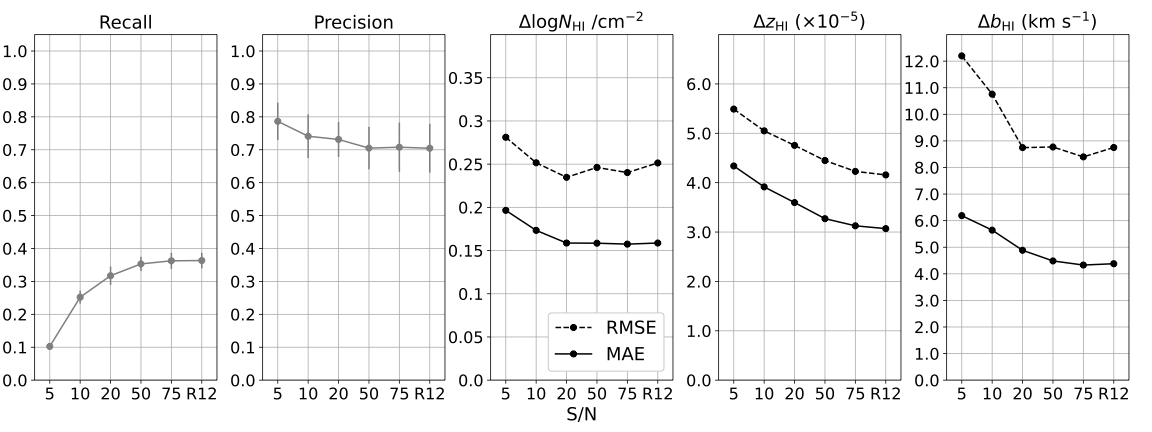}
   	\caption{CNN analysis of R12 spectra using a Gaussian distribution of S/N with a mean $=90$ and a standard deviation $=30$ (closer to the distribution of R12 spectra). From left to right, we present the recall, precision, the RMSE (black dashed line) and MAE (black solid line) of \HI\ column density ($ \Delta \log{N_{\mathrm{HI}}}/{\rm cm}^{-2}$), redshift ($\Delta z_{\mathrm{HI}}$), and the Doppler width ($\Delta b_{\mathrm{HI}}$).}
    \label{figapp:diffSNR_highSNR}
\end{center}
\end{figure*}
\begin{figure*}{}
\begin{center}
\graphicspath{}
	\includegraphics[width=2.1\columnwidth]{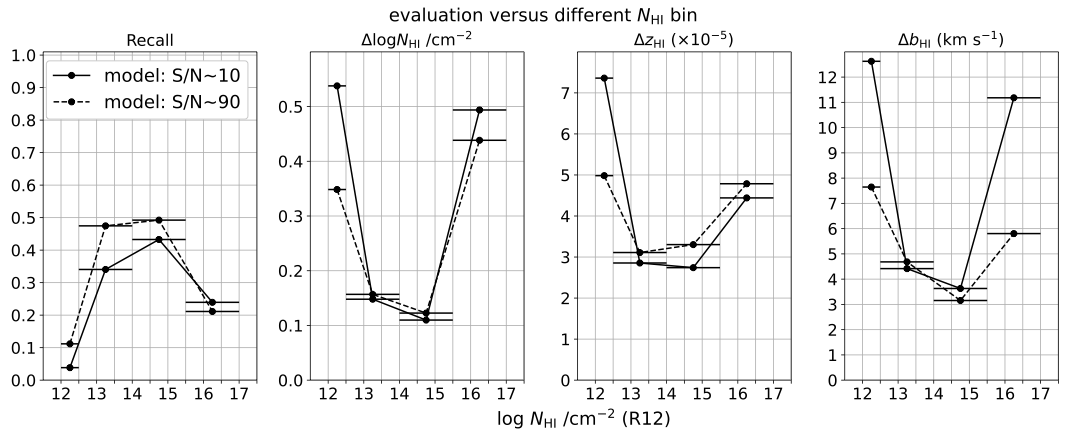}
   	\caption{Evaluation graphs of recall and the MAE of $\log{N_{\mathrm{HI}}}/{\rm cm}^{-2}$, redshift ($z_{\mathrm{HI}}$), and $b_{\mathrm{HI}}$ values, within different $\log{N_{\mathrm{HI}}}/{\rm cm}^{-2}$ bins. The horizontal lines of each datapoint represent the range of each bin; the intervals are 0.5, 1.5, 1.5, and 1.5, from low to high values of $\log{N_{\mathrm{HI}}}/{\rm cm}^{-2}$. The solid line shows the results using models trained by lower S/N (mean$=10$) while the dashed line shows the ones using models trained by higher S/N (mean$=90$).}
    \label{figapp:diffNHI_diffSNRmodel}
\end{center}
\end{figure*}
\begin{table}
	\centering
	\begin{tabular}{llc}
		\hline
		\multicolumn{1}{c}{} & {Hyperparameters} & {Optimised value}\\
		\hline\hline
		\multicolumn{1}{l}{Data Input} & {window size ($ws$)} & {285} \\
		\multicolumn{1}{l}{} & {$cnpix$} & {1} \\
		\hline
		\multicolumn{1}{l}{CNN} & {L2} & {0.0} \\
		\multicolumn{1}{l}{Architecture} & {dropout} & {0.6} \\
		\multicolumn{1}{l}{} & {conv\_filter\_1} & {256} \\
		\multicolumn{1}{l}{} & {conv\_filter\_2} & {256} \\
		\multicolumn{1}{l}{} & {conv\_filter\_3} & {512} \\
		\multicolumn{1}{l}{} & {conv\_kernel\_1} & {10} \\
		\multicolumn{1}{l}{} & {conv\_kernel\_2} & {7} \\
		\multicolumn{1}{l}{} & {conv\_kernel\_3} & {6} \\
		\multicolumn{1}{l}{} & {dense\_1} & {64} \\
		\multicolumn{1}{l}{} & {dense\_2\_ID} & {512} \\
		\multicolumn{1}{l}{} & {dense\_2\_N} & {512} \\
		\multicolumn{1}{l}{} & {dense\_2\_z} & {256} \\
		\multicolumn{1}{l}{} & {dense\_2\_b} & {128} \\
		\hline
	\end{tabular}
	\caption{Hyperparameters used in a CNN architecture trained using spectra with a higher S/N. These values are selected using a Bayesian optimisation algorithm.}
	\label{apptab:hyper-parameters_highSNR}
\end{table}
Extending the discussion in Section~\ref{app:diff_snr}, we have trained a CNN model with spectra of a higher S/N than the one used in the main work, and tested this model with the observed spectra from R12. The S/N of the training data is drawn from a Gaussian distribution of S/N with a mean $=90$ and a standard deviation $=30$; these values are chosen to be close to the S/N of the R12 data. The optimised hyperparameters of this CNN architecture are listed in Table~\ref{apptab:hyper-parameters_highSNR}. Comparing Fig.~\ref{figapp:diffSNR_highSNR} with Fig.~\ref{fig:HIRES_diffSNR}, the recall increases slightly but the performance drops when S/N $<50$. However, Fig.~\ref{figapp:diffNHI_diffSNRmodel} shows that by training a model with high S/N spectra, it helps to improve the predictions of systems within the range of lower ($\log{N_{\mathrm{HI}}}/{\rm cm}^{-2}<12.5$) and higher ($\log{N_{\mathrm{HI}}}/{\rm cm}^{-2}\geq15.5$) column density.


\label{lastpage}
\end{document}